\documentclass[12pt]{article}

\usepackage{newtxtext,newtxmath}

\usepackage{graphicx}

\usepackage[letterpaper,margin=1in]{geometry}

\renewenvironment{abstract}
	{\quotation}
	{\endquotation}

\date{}

\makeatletter
\renewcommand{\fnum@figure}{\textbf{Figure \thefigure}}
\renewcommand{\fnum@table}{\textbf{Table \thetable}}
\makeatother

\usepackage{scicite}

\usepackage{url}

\def\scititle{
	Structural reconstruction as the origin of the cuprate pseudogap
}
\title{\bfseries \boldmath \scititle}

\author{
	Sophie Beck$^{\ast}$ and 
	Aline Ramires$^{\ast}$\and
	Institute of Solid State Physics, TU Wien, 1040 Vienna, Austria.\and
	\small$^\ast$Corresponding authors. Emails: sophie.beck@tuwien.ac.at,  aline.ramires@tuwien.ac.at\and
}

\begin{document} 

\maketitle

\begin{abstract} \bfseries \boldmath




High-temperature superconductivity in the cuprates emerges from an enigmatic metallic state, known as the pseudogap, characterized by a reconstructed Fermi surface, reduced carrier density, and the appearance of Fermi arcs, whose origin remains unresolved. 
Here, we show that these defining signatures naturally arise from a structural reconstruction observed experimentally that introduces a symmetry-enforced sublattice degree of freedom.
In the presence of spin-orbit coupling, the Fermi surface is reconstructed into small closed pockets, effectively reducing the carrier density. 
The same sublattice structure gives rise to matrix-element interference in angle-resolved photoemission spectroscopy, leading to the manifestation of Fermi arcs. Density functional theory calculations support this mechanism. 
These results demonstrate that lattice symmetry provides a unifying and experimentally verifiable framework for understanding the pseudogap regime in the cuprates.
\end{abstract}

\noindent
The class of materials known as cuprates achieves the highest superconducting transition temperatures at ambient pressure, making them of fundamental and practical interest.
Yet, despite decades of research, a coherent understanding of the electronic phases that precede superconductivity in these materials remains elusive, posing a major obstacle to a microscopic theory of high-temperature superconductivity. 
In a broad region of the temperature–doping phase diagram, an anomalous metallic state, known as the pseudogap \cite{Timusk1999, Keimer2015, Proust2019}, is characterized by a reconstructed Fermi surface \cite{Sebastian2012, Fang2022}, a pronounced reduction in the number of mobile charge carriers \cite{Badoux2016, Collignon2017, Michon2018, Putzke2021}, and the appearance of Fermi arcs in angle-resolved photoemission spectroscopy (ARPES) experiments \cite{Norman1998, Vishik2018, Damascelli2003}. 
These features reflect a substantial reorganization of the electronic structure and are widely regarded as central to understanding the emergence of superconductivity in the cuprates. 

Although a variety of symmetry-breaking electronic orders, including charge- and spin-density waves, have been experimentally identified in parts of the phase diagram associated with the pseudogap, they emerge over a more restricted range of temperatures or dopings \cite{Comin2016, Frano2020}, suggesting that these symmetry-breaking order parameters are unlikely to constitute the primary origin of pseudogap behavior. 
Theoretically, the pseudogap has been approached from a wide range of perspectives. 
Proposals have invoked correlations inherited from the antiferromagnetic parent compound \cite{Anderson1987, Yang2006, Senthil2005}, superconducting fluctuations \cite{Emery1995}, hidden or intra–unit-cell orders \cite{Varma1997}, intertwined orders \cite{Fradkin2015}, and fractionalized excitations \cite{Sachdev2018}. 
While each of these ideas addresses selected aspects of the phenomenology, a longstanding difficulty has been to simultaneously account for small Fermi pockets \cite{Sebastian2012, Fang2022} and the concomitant appearance of Fermi arcs \cite{Norman1998, Vishik2018, Damascelli2003} within a single microscopic framework. This persistent inconsistency suggests that an essential ingredient in modeling the normal-state electronic structure of the cuprates may be missing.

Despite substantial variations in chemical composition and crystal structure, materials across different cuprate families exhibit strikingly similar phase diagrams, including the emergence of a pseudogap \cite{Timusk1999, Keimer2015, Proust2019}. 
This apparent universality has naturally guided the community to seek a minimal theoretical description capable of capturing the anomalous metallic behavior common to all cuprates.
A defining feature shared by all members of this class of materials is the strong two-dimensional character of the mobile charge carriers, which predominantly reside in the Cu-O$_2$ planes.
This observation has suggested that abstracting away material-specific structural details might nevertheless retain the essential physics underlying the universal phase diagram.
As a result, much of the theoretical effort has focused on simple models defined on a two-dimensional square lattice subject to strong electronic interactions, namely, on the Hubbard model.
A wide range of powerful analytical and numerical techniques has consequently been developed and employed yielding important insights into correlation-driven phenomena in the Hubbard model \cite{Qin2022}.
However, despite significant theoretical effort and extensive progress, reproducing all defining experimental signatures of the pseudogap within such minimal frameworks has remained challenging.

In contrast to the minimal two-dimensional descriptions, here we show that the explicit accounting of crystallographic complexity in the cuprates is key, as it naturally captures the main defining properties of the pseudogap.
We use as a guiding exxample doped La$_2$CuO$_4$, one of the cuprates best characterized from both crystallographic and electronic perspectives. 
In this material, as the system transitions from a high-temperature tetragonal (HTT) to a low-temperature orthorhombic (LTO) phase, the lattice acquires an intrinsic sublattice degree of freedom, with two sublattices related by a nonsymmorphic glide symmetry (as shown in the right and left panels of Fig.~\ref{fig:structural}). 
We find that this symmetry-enforced sublattice structure reconstructs the electronic bands near the Fermi level, which, in the presence of spin–orbit coupling (SOC), hybridize, leading to small closed Fermi surface pockets together with a reduction in the number of mobile carriers. 
The same sublattice degree of freedom gives rise to matrix-element interference in ARPES experiments, reconciling the well-established appearance of Fermi arcs with an underlying pocket-like Fermi surface. 
Using a combination of symmetry analysis and density functional theory (DFT) calculations, we demonstrate that this structural mechanism qualitatively and quantitatively captures the defining experimental signatures of the pseudogap.

\subsection*{A structure-property relation}

In the temperature-doping (T-p) phase diagram, the pseudogap phenomenology manifests below a doping-dependent temperature, usually referred to as $T^*$, and up to a maximum doping value $p^*$, as schematically depicted in Fig.~\ref{fig:structural} (center). Remarkably, for some cuprates, it has been established that a structural phase transition, characterized by a doping-dependent transition temperature $T_S$, happens near $T^*$.
In particular, structural studies of doped La$_2$CuO$_4$ have consistently reported a line of phase transitions between an HTT and an LTO phase ending at zero temperature close to $p^*$ \cite{Fleming1987, Axe1989, CyrChoinire2018}, which we identify with the dashed line in Fig.~\ref{fig:structural} (center). 
Furthermore, recent angle-dependent magnetoresistance measurements in La$_{1.6-x}$Nd$_{0.4}$Sr$_x$CuO$_4$ provided exquisite Fermi surface detail around $p^*$, revealing small closed Fermi surface pockets in the pseudogap regime, and a large Fermi surface sheet outside the pseudogap region \cite{Fang2022}. Drawing on these observations, we propose a structure-property relation that accounts for the most salient features of the pseudogap in the cuprates.

\subsection*{Increasing crystallographic complexity}

\begin{figure} 
	\centering
	\includegraphics[width=0.9\textwidth]{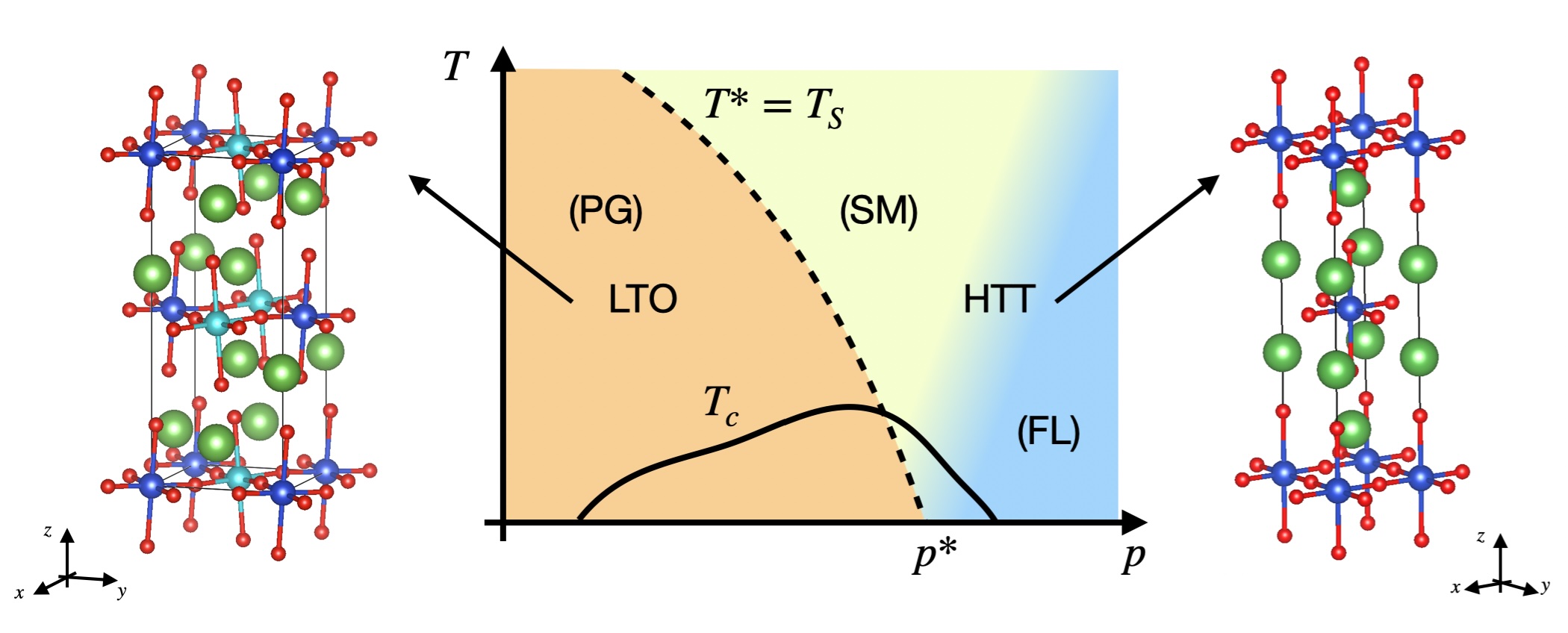} 

	\caption{
    \textbf{Structure-property relation in the pseudogap.} Center: Schematic temperature-doping ($T-p$) phase diagram for cuprate superconductors highlighting the structural origin of the pseudogap (PG). The dotted line marks both the onset of the PG, usually referred to as $T^*$, and the structural transition from the HTT to the LTO phase, $T_S$. In the HTT phase, both Fermi-liquid (FL) and strange-metallic (SM) behavior are reported. In the LTO phase, PG behavior is observed. The full line schematically marks the onset of superconductivity at the critical temperature $T_c$. Left and Right: 3D view of the crystal structure of the LTO and HTT phases of La$_2$CuO$_4$, respectively. La atoms are depicted as large green spheres, O as small red spheres, and Cu as medium blue spheres. Note the two shades of blue in the LTO phase, highlighting the two Cu sublattices. The thin line marks the conventional unit cell with four and two Cu atoms, respectively. Side panels produced by VESTA \cite{VESTA}.
    }
	\label{fig:structural} 
\end{figure}

\begin{figure} 
	\centering
	\includegraphics[width=0.9\textwidth]{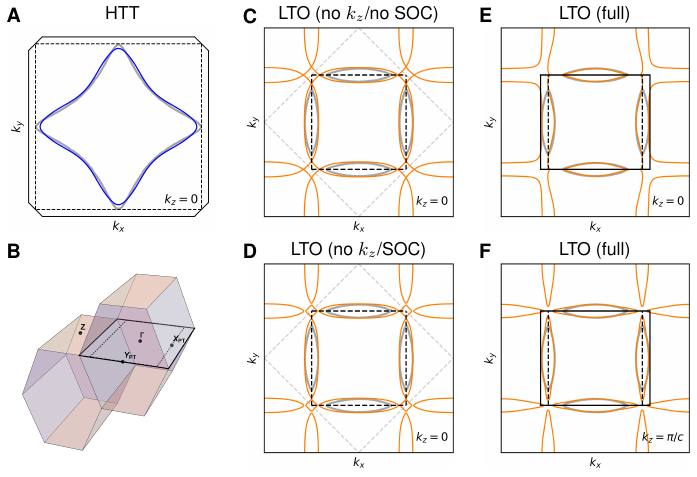} 
	\caption{
   \textbf{The pseudogap as a structural transition for doped La$_2$CuO$_4$.}
  (\textbf{A}) Fermi surface of the HTT phase for $p=0.24$ in the $k_z=0$ plane obtained by DFT (full blue line). The full black line indicates the edges of the 3D BZ and the dashed black lines the planar BZ.
  (\textbf{B}) 3D view of the BZ in the LTO phase, with face-centered orthorhombic character. We highlight the $k_z=0$ boundary of the BZ with a thick full line, the PTBZ with a thick dashed line, and the high-symmetry points $\Gamma$, $X_{PT}$, and $Y_{PT}$. 
  (\textbf{C}) Fermi surface of the LTO phase for $p=0.18$ in the $k_z=0$ plane for a DFT-derived tight-binding model without $k_z$-dependent terms or SOC (full orange line).
  The dashed black line marks the PTBZ, and the dashed gray line marks the planar BZ of the HTT phase rotated by $45^\circ$ to highlight the band folding. 
  (\textbf{D}) Same as panel (\textbf{C}), but with SOC.
  (\textbf{E}) Fermi surface as predicted by DFT including SOC.
  (\textbf{F}) Same as panel (\textbf{E}), but for $k_z=\pi/c$.
  Experimental reference data in gray digitized from Ref.~\cite{Fang2022}.
    }
	\label{fig:case_study}
\end{figure}

We first consider the overdoped regime, above $p^*$, where La$_2$CuO$_4$ remains in the HTT phase down to the lowest temperatures. In this crystallographic phase, the space group is I4/mmm ($\#$139) with a single Cu atom in the primitive unit cell, see Fig.~\ref{fig:structural} (right).
The normal state in this regime is described by a single large Fermi surface, which is well captured by DFT calculations (see materials and methods). This is illustrated in Fig.~\ref{fig:case_study}A, in which we overlay the calculated Fermi surface for $p=0.24$ with the one reported by recent angle-dependent magnetoresistance experiments \cite{Fang2022}. These first-principles results can be parametrized by a tight-binding model with a single Cu atom per unit cell, see supplementary text and Figs.~\ref{fig:structure_HTT} and \ref{fig:fig_sm_htt}. 

For doping values $p<p^*$, at high temperatures, La$_2$CuO$_4$ also stabilizes in the HTT phase. However, below a certain temperature, a structural phase transition to the LTO phase with space group Bmeb ($\#$64) takes place.
The main difference between the HTT and LTO phases is the tilt of the oxygen octahedra, effectively doubling the size of the unit cell. In the HTT phase, the Cu-apical O bonds lie along the z-axis, whereas in the LTO phase, they are tilted along one specific Cu-O plaquette diagonal in a staggered fashion. This tilt gives rise to two inequivalent Cu sites in the primitive unit cell, as indicated by the light and dark blue colors in Fig.~\ref{fig:structural} (left), and introduces glide planes, characterizing the space group as nonsymmorphic. 

The presence of a sublattice structure in the LTO phase makes the electronic structure unavoidably more complex, as a tight-binding Hamiltonian needs to be encoded in a two-dimensional space of internal degrees of freedom. The tight-binding Hamiltonian in matrix form reads
\begin{equation}\label{Eq:H_LTO}
    H_{SL}(\mathbf{k}) = \sum_{a=0}^3 h_a(\mathbf{k}) \hat{\tau}_a,
\end{equation}
where $\hat{\tau}_i$ are Pauli matrices with $i=\{1,2,3\}$ and $\hat{\tau}_0$ is the two-dimensional identity matrix acting on sublattice space. 
The term with $a=0$ captures the intra-sublattice hopping processes that are identical for both sublattices, the term with $a=1$ captures inter-sublattice hopping processes, and the term with $a=3$ captures the asymmetry in certain intra-sublattice processes. 
In the presence of inversion and time-reversal symmetries, the term with $a=2$ is not symmetry-allowed. 
Note that this Hamiltonian has two eigenvalues, $E^\pm_{SL}(\boldsymbol{k}) = h_0(\boldsymbol{k}) \pm \sqrt{h_1^2(\boldsymbol{k})+h_3^2(\boldsymbol{k})}$, which are associated with two bands in the absence of any extra symmetry-breaking order parameter. 
The symmetries and specific functional form of the lowest-order tight-binding terms contributing to each $h_{a}(\mathbf{k})$ are given in the supplementary text.

The Fermi surfaces obtained by DFT calculations for $p=0.18$, simplified by taking into account only in-plane hopping processes, are shown in Fig.~\ref{fig:case_study}C. 
In this case, the Fermi surface is equivalent to the combination of the original and folded HTT Fermi surfaces along the pseudotetragonal (PT) Brillouin Zone (BZ). 
Including out-of-plane processes, a three-dimensional description of the BZ is necessary, taking into account the face-centered orthorhombic nature of the crystal. 
The BZ in the $k_z=0$ plane is rectangular, and the BZs are stacked in a staggered fashion in the $k_xk_z$-plane, see Fig.~\ref{fig:case_study}B.
Once $k_z$-dependent terms are introduced, the band crossings at the boundary of the PTBZ boundary segment containing the $X_{PT}$ point are lifted for all planes with $k_z\neq n \pi/c$  (where $n$ is an integer), while the ones around the segment containing the $Y_{PT}$ point are preserved by a glide symmetry ($X_{PT}$ and $Y_{PT}$ points defined in Fig.~\ref{fig:case_study}B), as illustrated in Fig.~\ref{fig:fig_sm_lto}A-B.

\subsection*{The role of spin-orbit coupling}

We now introduce a second ingredient that has received little attention thus far in the context of cuprates, namely, SOC. While thought to be small in the cuprates, experimental signatures of a finite SOC have been reported  \cite{Gotlieb2018, Barrera2025}. 
Explicitly including the spin degree of freedom, the Hamiltonian describing the LTO phase acquires further complexity, as it is necessarily encoded in a four-dimensional space intertwining sublattice and spin degrees of freedom:
\begin{equation}\label{Eq:H_LTO_SOC}
    H_{SL+SOC}(\mathbf{k}) = \sum_{\{a,b\}=0}^3 h_{ab}(\mathbf{k}) \hat{\tau}_a\otimes \hat{\sigma}_b,
\end{equation}
where $\hat{\sigma}_i$ are Pauli matrices with $i=\{1,2,3\}$, and $\hat{\sigma}_0$ is the two-dimensional identity matrix acting on spin space. 
In the presence of inversion and time-reversal symmetries, only six pairs of $(a,b)$ indices are symmetry-allowed. 
There are three spin-independent terms with $a=\{0,1,3\}$ and $b=0$, corresponding to the hopping processes described above for the spinless Hamiltonian in Eq. \ref{Eq:H_LTO}. 
In addition, there are three spin-dependent terms associated with even-momentum SOC with $a=2$ and $b=\{1,2,3\}$. 
DFT calculations estimate the SOC near the Fermi surfaces to be approximately 5-10 meV (see supplementary text).
As SOC does not break inversion or time-reversal symmetries, the band structure remains at least doubly degenerate across the entire BZ.
The Hamiltonian has two (doubly-degenerate) eigenvalues $E^\pm_{SL+SOC}(\boldsymbol{k}) = h_{00}(\boldsymbol{k}) \pm \sqrt{h_{10}^2(\boldsymbol{k})+h_{30}^2(\boldsymbol{k})+h_{s}^2(\boldsymbol{k})}$, where $h_{s}^2(\boldsymbol{k}) = \sum_{i=1,2,3}h_{2i}^2(\boldsymbol{k})$ captures the effect of SOC.
Note that the three SOC components are not concomitantly zero at any point in reciprocal space (except at the $Y_{PT}$ point) \cite{Park2019, Yang2014}, such that $h_{s}(\boldsymbol{k})$ is always finite, akin to the Kane-Mele term in graphene \cite{Kane2005} (see supplementary text for further discussion).

\begin{figure}
	\centering
	\includegraphics[width=0.9\textwidth]{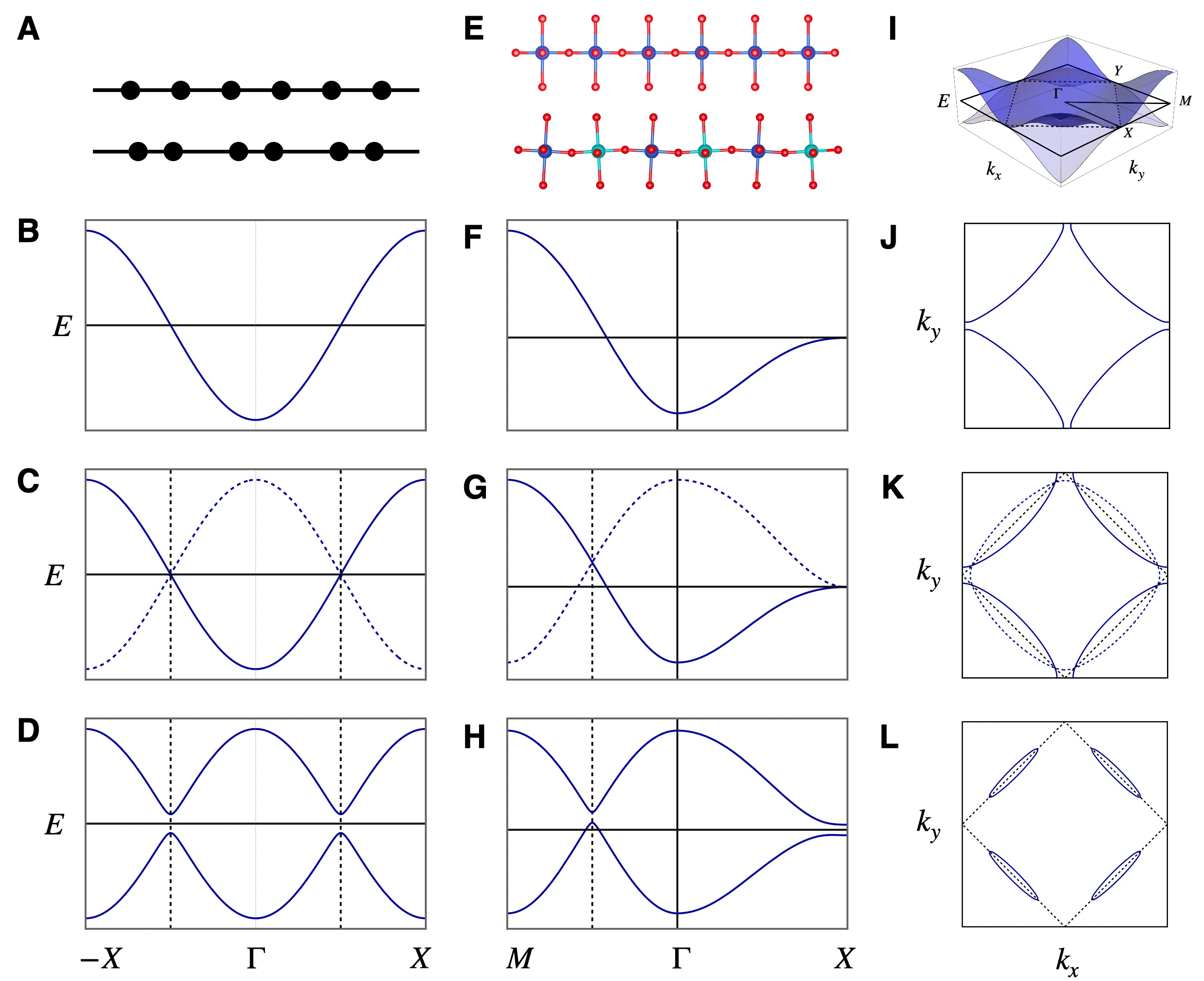} 
   \caption{\textbf{The pseudogap as a 2D Peierls instability.} 
   (\textbf{A}) Illustration of a monoatomic chain without (top) and with (bottom) dimerization.
   (\textbf{B}) Band structure of a monatomic chain with only nearest-neighbour hopping at half-filling along the 1D Brillouin zone (BZ). 
  (\textbf{C}) Same as panel (\textbf{B}) after an artificial zone folding corresponding to a unit cell doubling. The vertical dashed lines mark the new BZ boundary and the dashed line the folded band.
  (\textbf{D}) Same as panel (\textbf{B}) after dimerization.
  (\textbf{E}) Cu-O octahedra chains along the Cu-O bonds without (top) and with (bottom) the octahedra tilt. 
  (\textbf{F}) Schematic cuprate band structure close to $p^*$. 
  (\textbf{G}) Same as panel (\textbf{F}) after an artificial zone folding corresponding to a unit cell doubling. The vertical dashed lines mark the new BZ boundary and the dashed line the folded band.
  (\textbf{H}) same as panel (\textbf{F}) after octahedra tilt.
  (\textbf{I}) Schematic 2D view of the folded band structure of cuprates close to $p^*$. Highlighted are the high-symmetry points (with labels corresponding to the pre-folded BZ) and BZ boundaries before (full line) and after (dashed line) octahedra tilt. 
  (\textbf{J}-\textbf{L}) Fermi surfaces corresponding to bands in panels (\textbf{F}-\textbf{H}), respectively. Panel (\textbf{L}) highlights the formation of small Fermi pockets.  
  }
	\label{fig:peierls}
\end{figure}

Including SOC, in the absence of $k_z$-dependent terms, the band crossings around the PTBZ edges are all lifted, going from four- to two-fold degenerate, and small Fermi pockets are formed, as shown in Fig.~\ref{fig:case_study}D. 
Upon including $k_z$-dependent terms, the Fermi surfaces at the $k_z=0$ plane shown in Fig.~\ref{fig:case_study}E, correspond to two small Fermi pockets and two open Fermi sheets - which, however, have significant $k_z$-dependence, with four small Fermi pockets generally observed for finite $k_z$, as shown in Fig.~\ref{fig:case_study}F. 
Note that the SOC-induced band splitting is reminiscent of a two-dimensional Peierls instability \cite{Tang1988, Park2019} - a coordinated reconstruction of electrons and nuclei to minimize the energy of the entire system. 
Here, the analog of dimerization in the classical one-dimensional Peierls instability is the tilt of the oxygen octahedra, which enables SOC to gap out a large fraction of the electronic states, as schematically illustrated in Fig.~\ref{fig:peierls}.

\subsection*{Change in carrier density}

Assuming the pseudogap is associated with the structural transition from the HTT to the LTO phase, giving rise to small Fermi pockets in the presence of SOC, the change in carrier density can be naturally understood. 
The inter-sublattice hopping term, $h_{10}(\mathbf{k})$, is the dominant term in the Hamiltonian in the LTO phase. It is associated with bonding and antibonding states (see Fig.~\ref{fig:properties}A and B) formed by the degrees of freedom associated with the two sublattices. In the presence of SOC, the antibonding states become gapped. 
In this light, the structural transition naturally accounts for the change in carrier density from $1+p$ above $p^*$ (in the HTT phase) to $p$ below $p^*$ inside the pseudogap phase (in the LTO phase), as shown in Fig.~\ref{fig:properties}C. 
The intrinsic disorder in doped cuprates most likely blurs the structural phase transition, giving rise to an apparent smooth crossover between these two trends in carrier density.

\subsection*{Matrix-element interference}

\begin{figure}
	\centering
	\includegraphics[width=0.9\textwidth]{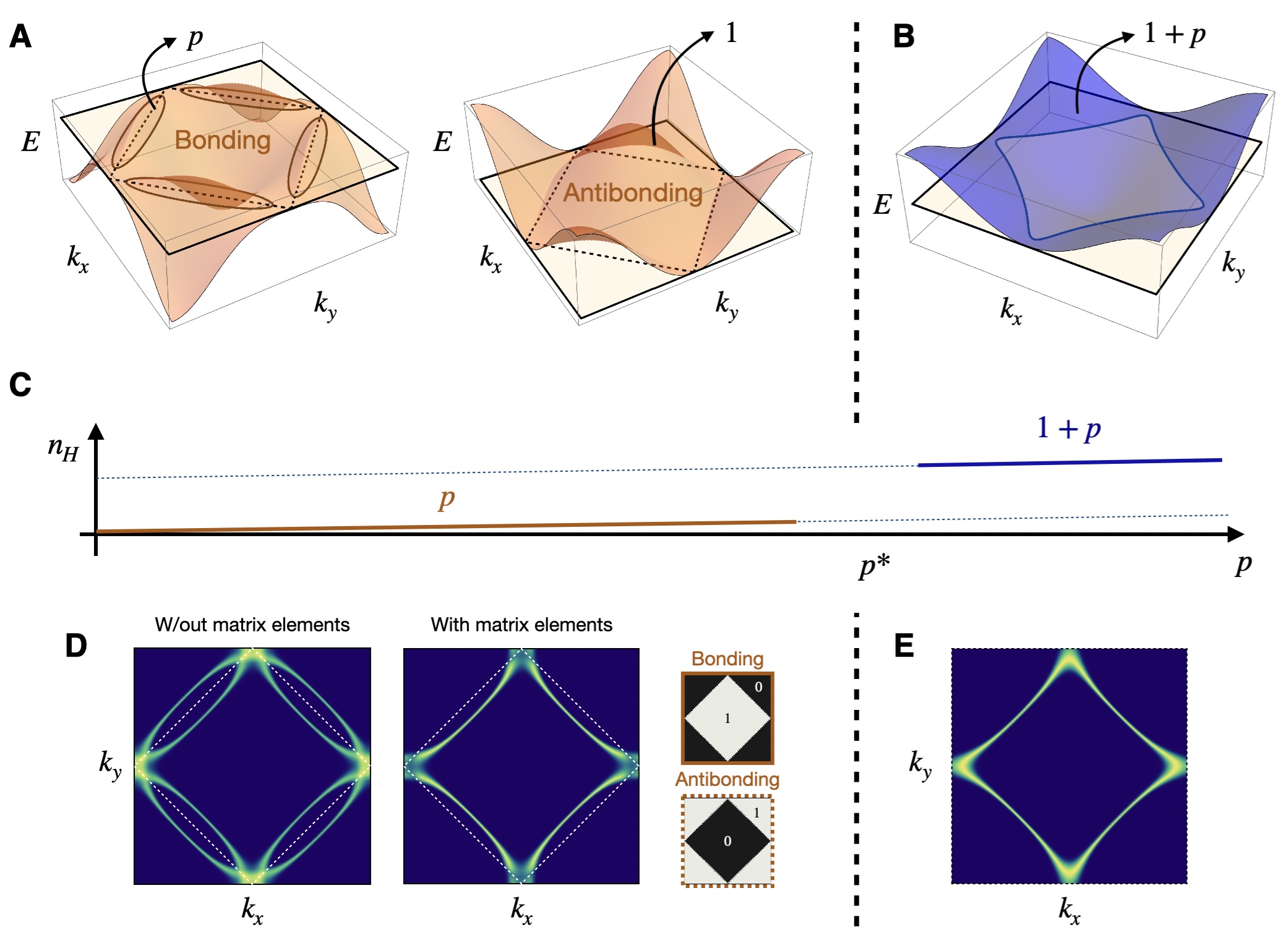} 
   \caption{\textbf{Main features of the pseudogap.} (\textbf{A}) Schematic bonding (left) and antibonding (right) bands in the LTO phase, plotted separately, excluding the $k_z$ dependence for a simpler discussion within the PTBZ (dashed lines) with enhanced SOC. Note that the antibonding band does not cross the Fermi energy (indicated by the light-coloured plane), and the bonding band is associated with small Fermi pockets highlighted by orange lines. 
  (\textbf{B}) Band structure in the HTT phase displaying a large Fermi surface highlighted by the blue line. 
  (\textbf{C}) Evolution of carrier density $n_H$ with doping $p$, assuming the structural phase transition at $p^*$.  
  (\textbf{D})  ARPES spectra (left) corresponding to the Fermi surface in panel (\textbf{A}) with and without matrix elements. The normalized matrix elements associated with the bonding and antibonding bands are shown in the right-most panels. 
  (\textbf{E}) ARPES spectrum corresponding to the Fermi surface in panel (\textbf{B}).
  In the ARPES spectra, dark blue corresponds to low intensity and bright yellow to maximum intensity. Parameters used for these plots are given in the supplemental information.
  }
	\label{fig:properties}
\end{figure}

A remaining puzzle associated with the pseudogap is the observation of Fermi arcs by ARPES experiments. 
Based on the scenario proposed here, the Fermi arcs can be understood from the presence of a sublattice structure in the LTO phase, which leads to destructive interference for the ARPES matrix elements \cite{Moser2017, Chung2024}. 
The matrix elements are dominated by the strongest term in the Hamiltonian, $h_{10}(\mathbf{k})$, associated with inter-sublattice hopping, and are approximately $\propto 1\pm h_{10}(\mathbf{k})/|h_{10}(\mathbf{k})|$ for the antibonding and bonding bands, respectively (see complete expressions in the supplementary information). 
In the absence of SOC, these form factors give rise to an ARPES spectrum reminiscent of a single large Fermi surface, as the matrix elements ``erase” one of the bands in each BZ, as shown in Figs.~\ref{Fig:ARPES_LTO_0} and \ref{Fig:ARPES_LTO_d}. 
This corresponds to a trivial band folding, which is effectively the case for such a band structure without $k_z$-dependent hopping processes or SOC encoding the fact that the two sublattices are actually distinct. 
In the presence of SOC, however, with the gapping of one of the Fermi surfaces, the matrix elements ``erase” the ARPES amplitude associated with the part of the Fermi pocket in the second PTBZ, giving rise to Fermi arcs, as shown in Fig.~\ref{fig:properties}D and Fig.~\ref{Fig:ARPES_LTO_d_SOC}.
This type of matrix-element effect, associated with a BZ selectivity, is well known in graphene \cite{Shirley1995}. Note that further terms in the Hamiltonian also contribute to the matrix elements, slightly deforming the form factors presented in Fig.~\ref{fig:properties}D, and potentially allowing for the detection of SOC through spin-resolved ARPES (explicit formulas for the spin-dependent matrix elements are given in the supplementary text).

\subsection*{Discussion and Outlook}

We have shown that a simple unified description of the pseudogap regime in the cuprates is achieved by explicitly accounting for crystallographic symmetry, including the nonsymmorphic nature of the low-temperature orthorhombic phase and the presence of spin–orbit coupling.
Within this framework, the defining phenomenology of the pseudogap emerges once the full structural complexity of the lattice is taken into account, without invoking additional electronic ordering phenomena.
The ability of this symmetry-based description to capture the reconstruction of the Fermi surface, the formation of small pockets, and the appearance of Fermi arcs, together with its quantitative agreement for doped La$_2$CuO$_4$, suggests an Occam’s-razor resolution of the pseudogap problem, effectively identifying the pseudogap \emph{phase} as a reconstructed Fermi liquid tied to a structural phase transition \cite{Mirzaei2013, Barisic2013}.

Although our symmetry analysis and DFT calculations focus on La$_2$CuO$_4$, the underlying mechanism is expected to be broadly relevant to other cuprate materials that exhibit comparable crystallographic complexity. Beyond the cuprates, our results naturally connect to reports of pseudogap-like behavior in materials such as Sr$_2$IrO$_4$ \cite{Kim2015, Peng2022, Alexanian2025} and Ca$_3$Ru$_2$O$_7$ \cite{Markovi2020}, in which crystallographic complexity and SOC have already been acknowledged to play important roles \cite{Kim2008}. More broadly, this perspective places the cuprates within a wider class of materials, including ferroelectrics and shape-memory alloys, in which displacive structural phase transitions are known to induce profound changes in electronic properties, with features reminiscent of a pseudogap \cite{Grunebohm2023}.

The structure–property relation identified here suggests natural connections to additional anomalies in the cuprate phase diagram. In particular, the linear-in-temperature resistivity observed above the critical doping associated with the pseudogap may arise from scattering by soft phonon modes near the structural transition between the HTT and LTO phases \cite{Birgeneau1987, Boni1988}. This interpretation highlights lattice dynamics as a potentially important ingredient in understanding transport anomalies in the normal state of the cuprates.

Finally, our results identify crystal symmetry as an experimentally tunable control parameter of the pseudogap phase. Uniaxial \cite{Gao2025, Guguchia2023, Thomarat2024} and epitaxial \cite{Ivashko2019} strain engineering provide direct routes to systematically manipulate lattice symmetry and test the proposed mechanism across different cuprate families. 
More broadly, our findings motivate a systematic exploration of how crystallographic symmetry influences superconductivity itself, including its interplay with electronic reconstruction and spin–orbit coupling.


\newpage


\clearpage 

%
\bibliography{biblio} 
\bibliographystyle{sciencemag}

%
%
%
%
%
%


\section*{Acknowledgments}
We thank N. Bari\v{s}i\'{c} and A. Hampel for discussions.

\paragraph*{Funding:}
S.B. acknowledges funding from the Simons Foundation (00010503, AT).

\paragraph*{Author contributions:}
S.B. performed the ab-initio calculations and A.R. the symmetry analysis. Both authors wrote the paper.

\paragraph*{Competing interests:}
There are no competing interests to declare.

\paragraph*{Data and materials availability:}
All information needed to reach the conclusions of this paper is presented in the main text and in the supplementary materials.
The ab initio input and output files, Wannier Hamiltonians, derived datasets, and analysis scripts that support the findings of this study have been deposited in Zenodo (DOI: doi.org/10.5281/zenodo.18836350), currently under embargo during peer review and scheduled for public release upon publication.
Additional information necessary to reproduce the calculations is available from the corresponding authors upon reasonable request.


\subsection*{Supplementary materials}
Materials and Methods\\
Supplementary Text\\
Figs. S1 to S8\\
References \textit{(50-\arabic{enumiv})\\ 
}


\newpage


\renewcommand{\thefigure}{S\arabic{figure}}
\renewcommand{\thetable}{S\arabic{table}}
\renewcommand{\theequation}{S\arabic{equation}}
\renewcommand{\thepage}{S\arabic{page}}
\setcounter{figure}{0}
\setcounter{table}{0}
\setcounter{equation}{0}
\setcounter{page}{1} 


\begin{center}
\section*{Supplementary Materials for\\ \scititle}

Sophie Beck$^{\ast}$ and 
	Aline Ramires$^{\ast}$\\ 
\small$^\ast$Corresponding authors. Emails: sophie.beck@tuwien.ac.at,  aline.ramires@tuwien.ac.at
\end{center}

\subsubsection*{This PDF file includes:}
Materials and Methods\\
Supplementary Text\\
Figures S1 to S8

\newpage


\subsection*{Materials and Methods}

First-principles calculations were performed within density functional theory (DFT) using the plane-wave pseudopotential method as implemented in Quantum ESPRESSO\cite{Giannozzi_et_al:2009}.
The exchange–correlation functional was treated within the generalized gradient approximation (GGA) using the Perdew-Burke-Ernzerhof (PBE) functional\cite{Perdew/Burke/Ernzerhof:1996}.
Core-valence interactions were described using ultrasoft pseudopotentials from the GBRV library\cite{Garrity_et_al:2014}.
A plane-wave kinetic energy cutoff of 55 Ry for the wavefunctions and 660 Ry for the charge density was employed.
Brillouin-zone integrations were carried out using a $\Gamma$-centered Monkhorst-Pack k-point mesh of 7$\times$7$\times$7, and all calculations were converged with respect to k-point sampling and energy cutoffs.

We consider two structural phases of La$_2$CuO$_4$: the low-temperature orthorhombic (LTO) phase and the high-temperature tetragonal (HTT) phase.
For the LTO structure, experimental lattice parameters and internal atomic positions reported for La$_{1.9}$Ba$_{0.1}$CuO$_4$ in Ref.\cite{Axe1994} were used to stabilize the orthorhombic distortion, while maintaining the stoichiometry of pure La$_2$CuO$_4$.
For the HTT phase, lattice parameters and internal atomic positions were fixed to experimentally reported values for La$_2$CuO$_4$\cite{BRADEN1994396}.
The Brillouin-zone sampling density was kept constant across both structures.
For the LTO structure we use a denser Brillouin-zone mesh of 11$\times$11$\times$13 to achieve reliable convergence in the presence of SOC, and apply a real-space cutoff of 30 $\mathring{A}$ to suppress long-range numerical noise.

Low-energy effective Hamiltonians were constructed using maximally localized Wannier functions as implemented in Wannier90\cite{Pizzi_2020}, interfaced with Quantum ESPRESSO.
Wannier functions were initialized from atomic-like projections  corresponding to the Cu $d_{x^2-y^2}$ orbitals that define the low-energy electronic structure, resulting in one and two maximally localized Wannier functions in the HTT and LTO phase, respectively.
The Wannier Hamiltonians accurately reproduce the low-energy band structure. 
These Hamiltonians were used to interpolate the electronic structure on arbitrary momentum grids and to analyze the minimal tight-binding models.
The tight-binding analysis was performed using the TRIQS software library~\cite{Parcollet_et_al:2015}, the solid\_dmft software package~\cite{Merkel_et_al:2022}, and the pythTB package~\cite{Coh/Vanderbilt:2022}.

To investigate the effect of spin–orbit coupling (SOC), we construct a Wannier Hamiltonian from a fully relativistic, noncollinear calculation (including SOC on Cu only) using norm-conserving pseudopotentials~\cite{VANSETTEN201839}, while keeping the Wannier projections and energy windows fixed.
We define an effective SOC contribution as the difference between the relativistic and scalar-relativistic Wannier Hamiltonians.
Within this construction, SOC manifests predominantly as on-site and short-range terms with magnitudes of approximately 1–6 meV, consistent with expectations for the Cu $3d$ spin–orbit interaction after projection onto a low-energy Wannier basis.
To vary the SOC strength shown in Fig. S4E-F, we construct an interpolated Hamiltonian by adding a scaled SOC contribution to the scalar-relativistic Hamiltonian.

In the vicinity of the Fermi surface, SOC lifts degeneracies along the Y–S direction, leading to band splitting on the order of 10 meV. 
We note that due to the gauge freedom inherent in Wannier representations, the precise matrix decomposition of the Hamiltonian into SOC and non-SOC contributions is not unique. However, the resulting band splittings are gauge-invariant and provide a robust measure of the SOC effects.

\subsection*{Supplementary Text}
This supplement provides details on the symmetry analysis and the construction of the effective tight-binding model referred to in the main text. We also provide details of spin-resolved ARPES matrix elements.

\subsubsection*{The high-temperature tetragonal phase (HTT)}

\begin{figure}
	\centering
	\includegraphics[width=0.9\textwidth]{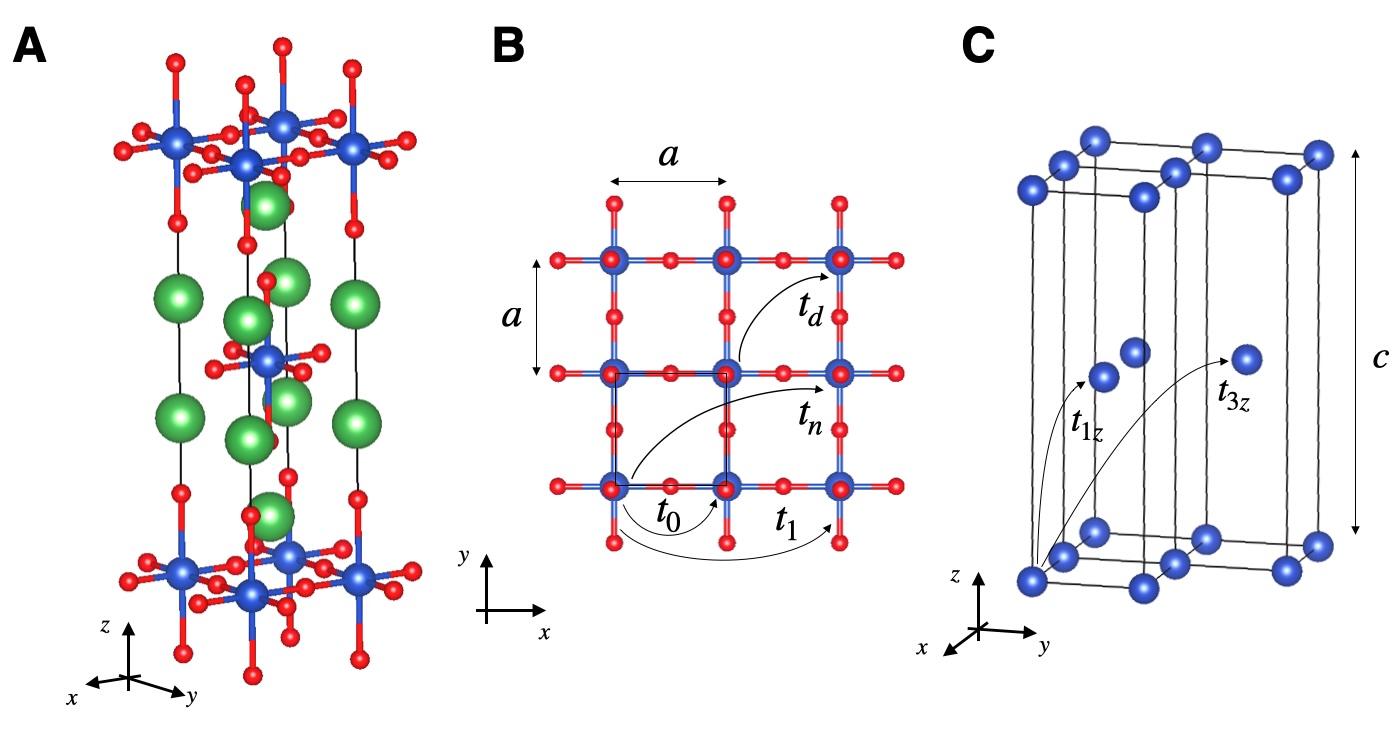} 
   \caption{\textbf{Structure and hopping parameters in the HTT phase.} (\textbf{A}) 3D view of the  HTT phase of La$_2$CuO$_4$ with La atoms as large green spheres, O as small red spheres, and Cu as medium blue spheres. The thin line marks the conventional unit cell with two equivalent Cu atoms. (\textbf{B}) Single plane of Cu-O octahedra indicating the lattice dimensions and the hopping processes considered here. (\textbf{C}) Simplified 3D view of the lattice with only Cu atoms indicating the lattice dimensions and hopping process along the $z$-direction. Figure made with VESTA \cite{VESTA}.
   }
	\label{fig:structure_HTT}
\end{figure}

In the HTT phase, La$_2$CuO$_4$ crystallizes with space group $I4/mmm$ ($\# 139$), a symmorphic group characterized by sixteen operations plus their combination with $\mathbf{\tau}_c=(1/2,1/2,1/2)$, a vector associated with the body-centered nature of the lattice.  
The ``I" label indicates the lattice is body-centered.
The ``4" label indicates there is a four-fold rotational axis along the $z$-axis.
The first ``m" label indicates that there is a mirror symmetry with normal along the $z$-direction.
The second ``m" label indicates the presence of mirrors with normals along the $x$- and $y$-directions, while the third ``m" label indicates the presence of mirrors with normals along the in-plane diagonals. 
The combination of mirror operations implies the space group contains inversion symmetry.
Note that there is only one Cu site in the primitive unit cell. Fig. ~\ref{fig:structure_HTT}A displays the conventional unit cell with two (equivalent) Cu atoms.

Considering the most relevant in-plane and out-of-plane hopping processes, as indicated in Fig.~\ref{fig:structure_HTT}B-C, the electronic band dispersion is given by:
\begin{eqnarray}\label{Eq:TB_HTT}
E(\mathbf{k}) &=& -\mu +2 t_0 [\cos (k_x a) + \cos (k_y a)] 
\\ \nonumber
&&+2t_1 [\cos(2k_x a)+\cos (2k_y a)] 
\\ \nonumber
&&+ 4 t_d \cos(k_xa)\cos(k_ya)
\\ \nonumber
&&+4 t_n[\cos(2k_xa)\cos(k_ya) + \cos(k_xa)\cos(2k_ya)]
\\ \nonumber
&&+8 t_{1z}[\cos(k_xa/2)\cos(k_ya/2)]\cos(k_zc/2)
\\ \nonumber
&&+8 t_{3z} [\cos(3k_xa/2)\cos(3k_ya/2)]\cos(k_zc/2)
\end{eqnarray}

The parameters extracted from DFT are $t_0 = -0.4428$ eV, $t_1=-0.0363$ eV,  $t_d = 0.0305$ eV, $t_n=-0.0266$, $t_{1z} = -0.0083$ eV, $t_{3z} = 0.0092$ eV, and $\mu$ is the doping-dependent chemical potential. The lattice parameters are $a=3.807 \mathring{A}$ and $c=13.204 \mathring{A}$~\cite{BRADEN1994396}.

\begin{figure}
	\centering
	\includegraphics[width=0.9\textwidth]{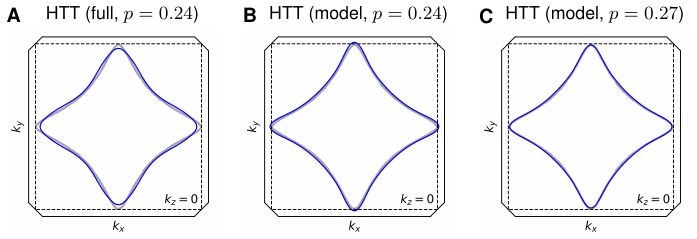} 
   \caption{\textbf{Fermi surface in the HTT phase.} 
   (\textbf{A}) Fermi surface obtained by DFT at $p=0.24$.
   (\textbf{B}) Fermi surface obtained by the tight-binding model above at $p=0.24$.
   (\textbf{C}) Same as (\textbf{B}), but for $p=0.27$.
   Fermi surface reported by Fang et al. \cite{Fang2022} is illustrated in gray lines.
   All Fermi surfaces are shown in the $k_z=0$ plane.
   }
	\label{fig:fig_sm_htt}
\end{figure}

We plot representative Fermi surfaces for the HTT phase in Fig.~\ref{fig:fig_sm_htt}. In panels (\textbf{A}) and (\textbf{B}), we plot the Fermi surfaces obtained from the full DFT calculations and from the tight-binding model above, respectively. We note that the apparent better match of the tight-binding model (compared with the full DFT solution) with the reported Fermi surface in Fang et al. \cite{Fang2022} is most likely associated with the fact that the parametrization of the Fermi surface based on experimental results relies on a model with few tight-binding parameters. For comparison, we also plot in panel (\textbf{C}) the Fermi surface obtained from the tight-binding model above with a slightly different doping. For each value of $p$, we use the corresponding rigid-band shift of the chemical potential, accounting for the change in particle number.

For Fig.~\ref{fig:properties}B in the main text, we use the tight-binding model in Eq.~\ref{Eq:TB_HTT}, with the parameters below it and $\mu=-0.29$ eV, corresponding to $p\approx 0.24$.

\subsubsection*{The low-temperature orthorhombic phase (LTO)}

In the LTO phase, La$_2$CuO$_4$ crystallizes with space group $Bmeb$ ($\# 64$), a nonsymmorphic space group characterized by eight operations plus their combinations with $\boldsymbol{\tau}_B=(1/2,0,1/2)$, a vector associated with the face-centered nature of the lattice. 
The ``B" label indicates that the centered atoms are on the surfaces with normal along the y-direction. 
The ``m" label indicates that the system has a simple mirror symmetry with normal along the $x$-direction. 
The ``e" label indicates that the system has a glide plane with normal along the $y$-direction, accompanied by the fractional translation vector $\boldsymbol{\tau}_e = (0,1/2,1/2)$ or $(1/2,1/2,0)$.
The ``b" label indicates there is a glide plane with normal along the z-direction with accompanying fractional translation vector $\boldsymbol{\tau}_b= (0,1/2,0)$ if the plane is between the Cu-O layers, or $\boldsymbol{\tau}_b= (0,1/2,1/2)$ if the plane coincides with the Cu-O layers.
The presence of the mirror and glide planes guarantees the presence of inversion symmetry. 
Inversion can be composed with the mirror and glide operations, generating a two-fold rotation axis and two screw axes.

\begin{figure}
	\centering
	\includegraphics[width=0.9\textwidth]{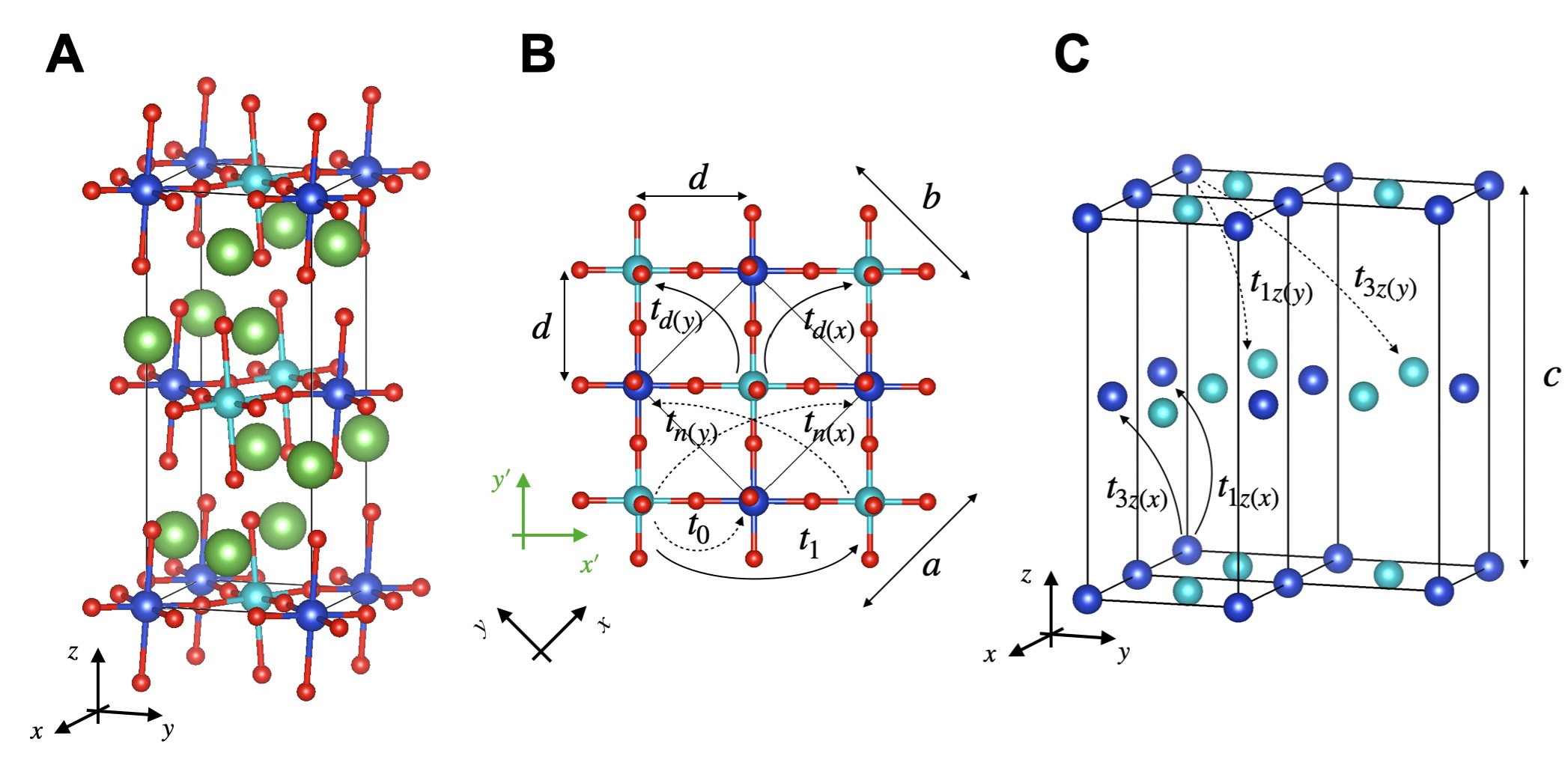} 
   \caption{\textbf{Structure and hopping parameters in the LTO phase.} (\textbf{A}) 3D view of the  LTO phase of La$_2$CuO$_4$ with La atoms as large green spheres and O as small red spheres. Note the two different types of Cu sites represented as medium dark/light blue spheres. The thin line marks the conventional unit cell with four atoms. (\textbf{B}) Single plane of Cu-O octahedra indicating the lattice dimensions and the hopping processes. The intra-sublattice hopping processes are indicated with a full line arrow, while the inter-sublattice hopping processes are indicated with a dashed line arrow. The thin full line indicates the 2D unit cell. Note the different coordinate system with respect to the HTT phase. The primed (green) coordinate system is the one used for the HTT phase, while the unprimed (black) coordinate system is the one we adopt for the LTO phase.
(\textbf{C}) Simplified 3D view of the lattice with only Cu atoms indicating the dimension and hopping processes along the $z$-direction.  Figure made with VESTA \cite{VESTA}.}
	\label{fig:structure_LTO}
\end{figure}

The main difference between the HTT and LTO phases is the tilt of the oxygen octahedra. In the HTT phase, the Cu-apical O bonds lie along the $z$-axis, whereas in the LTO phase, they are tilted along one specific Cu-O plaquette diagonal. This tilt gives origin to two inequivalent Cu sites in the primitive unit cell, or four Cu sites in the conventional unit cell, as indicated by the light and dark blue colors in Fig.~\ref{fig:structure_LTO}A.

We associate the two Cu sites with a sublattice structure, which we encode in a two-dimensional space of internal degrees of freedom in terms of Pauli matrices $\hat{\tau}_i$ ($i=1,2,3$) and the identity matrix $\hat{\tau}_0$, such that the Hamiltonian can be parametrized as:
\begin{eqnarray}
H_{SL}(\mathbf{k}) &=& \sum_{a=0}^3h_a(\mathbf{k})\hat{\tau}_a,
\end{eqnarray}
where $h_a(\mathbf{k})$ are functions of momentum. Given the Hermiticity of the Pauli matrices, the functions $h_a(\mathbf{k})$ must be real. Furthermore,  as inversion symmetry does not exchange sublattices (implemented as $\hat{\tau}_0$ and $\mathbf{k} \rightarrow -\mathbf{k}$), the functions $h_a(\mathbf{k})$ must be $\mathbf{k}$-even. In the presence of time-reversal symmetry (implemented as $\hat{\tau}_0$, $\mathbf{k} \rightarrow -\mathbf{k}$, followed by complex conjugation), $h_2(\mathbf{k})$ must be $\mathbf{k}$-odd, contradicting the previous condition, therefore $h_2(\mathbf{k})=0$. There is no symmetry condition indicating that any of the other $h_a(\mathbf{k})$ functions must be zero over the entire Brillouin Zone (BZ). Physically, $h_0(\mathbf{k})$ corresponds to the hopping processes that are the same for both sublattices, $h_1(\mathbf{k})$ corresponds to the hopping processes between sublattices, and $h_3(\mathbf{k})$ corresponds to an imbalance in the intra-sublattice hopping processes.

Taking into account the largest hopping processes, in analogy to the processes considered for the HTT phase, we find:
\begin{eqnarray}\label{Eq:TB_LTO}
h_0(\mathbf{k})  &=& -\mu +4t_1 \cos(k_x a)\cos(k_y b)
\\ \nonumber
&&+2t_{d(x)} \cos(k_x a) +2t_{d(y)}\cos(k_y b)
\\  \nonumber
&&+4 t_{1z(x)} \cos(k_x a/2)\cos(k_z c/2) 
\\
\nonumber
&&+4 t_{3z(x)}\cos(3k_x a/2)\cos(k_z c/2),
\\ \nonumber
h_1(\mathbf{k}) &=& 4 t_0\cos(k_x a/2)\cos(k_y b/2) \\ \nonumber
&&+4 t_{n(x)}\cos(3 k_x a/2)\cos( k_yb/2) + 4t_{n(y)} \cos(k_x a/2)\cos(3k_yb/2)
\\ \nonumber
 &&+4 t_{1z(y)} \cos(k_y b/2)\cos(k_z c/2)\\ \nonumber
 &&+t_{3z(y)} \cos(3 k_y b/2) \cos(k_z c/2).
\end{eqnarray}
 
Here $t_0 = -0.4532$ eV, $t_1=-0.0393$ eV, $t_{d(x)} = 0.0333$ eV,  $t_{d(y)} = 0.0279$ eV, $t_{n(x)} = -0.0260$ eV, and $t_{n(y)} = -0.0264$ eV, $t_{1z(x)}= -0.0082$ eV,   $t_{1z(y)}= -0.0080$ eV, $t_{3z(x)}= 0.0075$ eV, and  $t_{3z(y)}= 0.0069$ eV. The lattice parameters are  $a= 5.3436 \mathring{A}$, $b=5.3773 \mathring{A}$, $d = 3.7900 \mathring{A}$, and $c= 13.2386 \mathring{A}$~\cite{Axe1994}.

The electronic bands in the presence of the sublattice structure are given by:
\begin{eqnarray}
    E^\pm_{SL}(\mathbf{k}) = h_{0}(\mathbf{k}) \pm \sqrt{h_{1}^2(\mathbf{k})+h_{3}^2(\mathbf{k})}.
\end{eqnarray}
Note that, within the hopping processes included above, we find $h_3(\mathbf{k})=0$. This term is, in fact, nonzero and arises from out-of-plane hopping processes to more distant neighbors.
 
Representative Fermi surfaces for the LTO phase are displayed in Fig.~\ref{fig:fig_sm_lto}A-B. Note that there are two bands giving rise to two Fermi surfaces stemming from the presence of the two Cu sublattices. Note that $k_z$-dependent terms split the Fermi surfaces across the PTBZ edge containing the $X_{PT}$ point, but the Fermi surface crossings along the PTBZ edge containing the $Y_{PT}$ point remain symmetry-protected. Note also that there is a strong flattening of the Fermi surface outside the PTBZ around the $X_{PT}$ point, which might facilitate the development of modulated orders in the LTO phase.

\subsubsection*{Symmetry aspects of the LTO phase}

Above, we argued why $h_2(\mathbf{k})$ must be zero based on the presence of inversion and time-reversal symmetries. In this subsection, we highlight the implications of the crystalline symmetries for the explicit properties of the remaining $h_a(\mathbf{k})$ functions. Here we focus on the three generators of the space group, $M_x$, $\bar{M}_y$, and $\bar{M}_z$. The action of these transformations in Cartesian coordinates and sublattice space is the following:
\begin{eqnarray}
    &&M_x: (x,y,z)\rightarrow (-x,y,z); \hat{\tau}_0,\\
    &&\bar{M}_y: (x,y,z)\rightarrow (x,-y+1/2,z+1/2); \hat{\tau}_1,\\
    &&\bar{M}_z: (x,y,z)\rightarrow (x,y+1/2,-z+1/2) ; \hat{\tau}_1,
\end{eqnarray}

Considering the first crystalline symmetry as the mirror reflection $M_x$, the invariance of the Hamiltonian under this symmetry implies:
\begin{eqnarray}
{M}_x H_{SL}(\mathbf{k}) {M}_x^{-1} = H_{SL}(\mathbf{k}) \Rightarrow \sum_a h_a(M_x \mathbf{k}) \hat{\tau}_a = \sum_a h_a(\mathbf{k}) \hat{\tau}_a \Rightarrow h_a(M_x \mathbf{k}) = h_a(\mathbf{k}),
\end{eqnarray}
such that all $h_a(\mathbf{k})$ functions must be even in $k_x$.

Considering now the glide symmetry $\bar{M}_y$, the invariance of the Hamiltonian under this symmetry implies:
\begin{eqnarray}
{\bar{M}}_y H_{SL}(\mathbf{k}) {\bar{M}}_y^{-1} = H_{SL}(\mathbf{k}) \Rightarrow \sum_a h_a(M_y \mathbf{k})\hat{\tau}_1 \hat{\tau}_a \hat{\tau}_1 =  \sum_a h_a(\mathbf{k}) \hat{\tau}_a \Rightarrow \pm h_a(M_y \mathbf{k}) = h_a(\mathbf{k}),
\end{eqnarray}
with the plus sign for $a=\{0,1\}$ and a minus sign for $a=3$. This implies that $h_a(\mathbf{k})$ functions with $a=\{0,1\}$ must be even in $k_y$, while the function with $a=3$ must be odd in $k_y$.

Analogously, the presence of the glide symmetry $\bar{M}_z$ (implemented by $\hat{\tau}_1$ and and $k_z\rightarrow -k_z$) implies:
\begin{eqnarray}
{\bar{M}}_z H_{SL}(\mathbf{k}) {\bar{M}}_z^{-1} = H_{SL}(\mathbf{k}) \Rightarrow \sum_a h_a(M_z \mathbf{k})\hat{\tau}_1 \hat{\tau}_a \hat{\tau}_1 =  \sum_a h_a(\mathbf{k}) \hat{\tau}_a \Rightarrow \pm h_a(M_z \mathbf{k}) = h_a(\mathbf{k}),
\end{eqnarray}
with the plus sign for $a=\{0,1\}$ and a minus sign for $a=3$. This implies that $h_a(\mathbf{k})$ functions with $a=\{0,1\}$ must be even in $k_z$, while the function with $a=3$ must be odd in $k_z$.

From the analysis above, we know that the lowest order terms in powers of $\mathbf{k}$ for $h_{0}(\mathbf{k})$ and $h_{1}(\mathbf{k})$ can only be proportional to $1$, $k_x^2$, $k_y^2$, $k_z^2$, with independent coefficients, as no symmetry constraints them. Conversely, the lowest order terms for $h_{3}(\mathbf{k})$ should be proportional to $k_yk_z$.

Considering now the periodicity of the Hamiltonian in reciprocal space by translations by reciprocal lattice vectors $\mathbf{G}_i$, explicitly given by:
\begin{eqnarray}
    &&\mathbf{G}_1 = (2\pi/a,0,2\pi/c),\\
    &&\mathbf{G}_2 = (2\pi/a,0,-2\pi/c),\\
    &&\mathbf{G}_3 = (0,2\pi/b,0),
\end{eqnarray}
we find
\begin{eqnarray}
H_{SL}(\mathbf{k}+\mathbf{G}_i) = h_0(\mathbf{k}) \hat{\tau}_0 + e^{i \mathbf{G}_i\cdot \mathbf{v}_{12}}h_1(\mathbf{k}) \hat{\tau}_1 + h_3(\mathbf{k}) \hat{\tau}_3,
\end{eqnarray}
where the term with $a=1$ picks up a non-trivial phase associated with the relative position of the two sublattices in real space, captured by $\mathbf{v}_{12} = (a/2,b/2,0)$ or $(0,b/2,c/2)$. Note that all the products $\mathbf{G}_i\cdot \mathbf{v}_{12}=\pi$, imposing $h_1(\mathbf{k}+\mathbf{G}_i) = -h_1(\mathbf{k})$. For points that are taken into themselves by a RLV, we find $h_1(\mathbf{k}) = 0$. Given the explicit form of the RLVs above, we conclude that $h_1(\mathbf{k})$ must be zero at the BZ plane determined by $(k_x,\pm \pi/b, k_z)$, and along the lines determined by $(\pm \pi/a, k_y, \pm \pi/c)$.

In the light of the symmetry arguments above, in terms of periodic functions, $h_0(\mathbf{k})$ has terms proportional to $\cos(n k_i a_i)$ and their products, for integer $n$ and $i=\{x,y,z\}$ ($a_i$ is the lattice constant along the respective direction). Given the periodicity with respect to $\mathbf{G}_{1,2}$ terms of the form $\cos(p k_x a)\cos(q k_z c)$ are also allowed with $p$ and $q$ half-integers. Conversely, given the momentum space constraints by RLV translations, $h_1(\mathbf{k})$ can only have terms proportional to $\cos(p k_x a)\cos(q k_y b)$, $\cos(p k_y b)\cos(q k_z c)$, or products of these, with $p$ and $q$ half-integers. Note that the terms in the tight-binding model above satisfy these conditions. 

Finally, $h_3(\mathbf{k})$ is also constrained by the mirror symmetries such that its lowest order contribution is of the form $\cos(m k_x a)\sin(n k_y b)\sin(p k_z c)$. Furthermore, the constraints imposed by translations require $n$ to be an integer, and $m$ and $p$ to be both integers or both half-integers. 

Note that the constraints above determine that both $h_1(\mathbf{k})$ and $h_3(\mathbf{k})$ are zero at the BZ edge plane determined by $(k_x, \pm \pi/b, k_z)$, guaranteeing a protected band crossing at this BZ edge after band folding. Note, however, that there is no such constraint for the planes determined by $(\pm \pi/a, k_y, k_z)$, associated with the complementary PTBZ edge. This implies that the bands can hybridize at this BZ edge, leading to the formation of Fermi pockets, as shown in Fig.~\ref{fig:fig_sm_lto}A-B.

\subsubsection*{The role of SOC in the LTO phase}

In this subsection, we introduce spin-orbit coupling (SOC) and discuss its consequences for the electronic structure of the LTO phase of La$_2$CuO$_4$. As introduced in the main text, the most general form of the Hamiltonian in the presence of SOC is:
\begin{eqnarray}
H_{SL+SOC}(\mathbf{k}) &=& \sum_{\{a,b\}=0}^3h_{ab}(\mathbf{k})\hat{\tau}_a\otimes \hat{\sigma}_b,
\end{eqnarray}
where $h_{ab}(\mathbf{k})$ are functions of momentum. Given the Hermiticity of the Pauli matrices, the functions $h_{ab}(\mathbf{k})$ must be real. Inversion symmetry does not exchange sublattices or affect the spins (implemented as $\hat{\tau}_0\otimes \hat{\sigma}_0$ and $\mathbf{k} \rightarrow -\mathbf{k}$), implying that all functions $h_{ab}(\mathbf{k})$ must be $\mathbf{k}$-even. In addition, in the presence of time-reversal symmetry (implemented as $i \hat{\tau}_0 \otimes \hat{\sigma}_2$, $\mathbf{k} \rightarrow -\mathbf{k}$, followed by complex conjugation), the only symmetry-allowed terms in the Hamiltonian have indexes $(a,b)= \{(0,0),(1,0),(3,0),(2,1),(2,2),(2,3)\}$. The first three spin-independent terms already existed in the spin independent case discussed above [$h_{00}(\mathbf{k})=h_{0}(\mathbf{k})$, $h_{10}(\mathbf{k})=h_{1}(\mathbf{k})$, and $h_{30}(\mathbf{k})=h_{3}(\mathbf{k})$]. The three new terms correspond to SOC.

We can determine the momentum dependence of the new SOC terms based on the generators of the space group, $M_x$, $\bar{M}_y$, and $\bar{M}_z$. Supplementing the action of these symmetry transformations on Cartesian coordinates and sublattices, their action on the spin degree of freedom is:
\begin{eqnarray}
    &&M_x: i\hat{\sigma}_1,\\
    &&\bar{M}_y: i\hat{\sigma}_2,\\
    &&\bar{M}_z: i\hat{\sigma}_3.
\end{eqnarray}

Considering first the mirror symmetry $M_x$, the invariance of the Hamiltonian under this symmetry implies:
\begin{eqnarray}
{M}_x H_{SL+SOC}(\mathbf{k}) M_x^{-1} &=& \hat{\tau}_0\otimes (i\hat{\sigma}_1) h_{ab}(M_x \mathbf{k})  \hat{\tau}_a\otimes \hat{\sigma}_b  \hat{\tau}_0\otimes (-i\hat{\sigma}_1) 
\\ \nonumber
&=&h_{ab}(M_x \mathbf{k}) \hat{\tau}_a \otimes \hat{\sigma}_1\hat{\sigma}_b \hat{\sigma}_1,
\end{eqnarray}
such that $h_{21}(\mathbf{k})$ is $k_x$-even and $h_{22}(\mathbf{k})$ and $h_{23}(\mathbf{k})$ are $k_x$-odd.

Considering first the glide symmetry $\bar{M}_y$, the invariance of the Hamiltonian under this symmetry implies:
\begin{eqnarray}
\bar{M}_y H_{SL+SOC}(\mathbf{k}) \bar{M}_y^{-1} &=& \hat{\tau}_1\otimes (i\hat{\sigma}_2) h_{ab}(M_y \mathbf{k})  \hat{\tau}_a\otimes \hat{\sigma}_b  \hat{\tau}_1\otimes (-i\hat{\sigma}_2) 
\\ \nonumber
&=&h_{ab}(M_y \mathbf{k}) \hat{\tau}_1\hat{\tau}_a\hat{\tau}_1 \otimes \hat{\sigma}_2\hat{\sigma}_b \hat{\sigma}_2,
\end{eqnarray}
such that $h_{21}(\mathbf{k})$ and $h_{23}(\mathbf{k})$ are $k_y$-even and $h_{22}(\mathbf{k})$ is $k_y$-odd.

Analogously, the invariance of the Hamiltonian under $\bar{M}_z$ symmetry implies:
\begin{eqnarray}
\bar{M}_z H_{SL+SOC}(\mathbf{k}) \bar{M}_z^{-1} &=& \hat{\tau}_1\otimes (i\hat{\sigma}_3) h_{ab}(M_z \mathbf{k})  \hat{\tau}_a\otimes \hat{\sigma}_b  \hat{\tau}_1\otimes (-i\hat{\sigma}_3) 
\\ \nonumber
&=&h_{ab}(M_z \mathbf{k}) \hat{\tau}_1\hat{\tau}_a\hat{\tau}_1 \otimes \hat{\sigma}_3\hat{\sigma}_b \hat{\sigma}_3,
\end{eqnarray}
such that $h_{21}(\mathbf{k})$ and $h_{22}(\mathbf{k})$ are $k_z$-even and $h_{23}(\mathbf{k})$ is $k_z$-odd.

From the analysis above, we know that the lowest order terms in powers of $\mathbf{k}$ for $h_{21}(\mathbf{k})$ can only be proportional to $1$, $k_x^2$, $k_y^2$, $k_z^2$, with independent coefficients, as no symmetry constraints them. Conversely, the lowest order terms for $h_{22}(\mathbf{k})$ and $h_{23}(\mathbf{k})$ should be proportional to $k_xk_y$ and $k_xk_z$, respectively.

In analogy to the $h_{10}(\mathbf{k})$ terms discussed above, as the SOC terms carry a $\hat{\tau}_2$ matrix, corresponding to an inter-sublattice process, all SOC terms are subject to the constraint imposed by RLV translations $h_{2b}(\mathbf{k}+\mathbf{G}_i) = -h_{2b}(\mathbf{k})$. This constraint imposes further restrictions on the $h_{2b}(\mathbf{k})$ in terms of periodic functions in reciprocal space. 

In particular, for $h_{21}(\mathbf{k})$ to satisfy this condition, it cannot have a constant term and the lowest order terms are
\begin{eqnarray}
    h_{21}(\mathbf{k}) = \lambda_{1a} \cos(k_y b/2)\cos(k_x a /2)+\lambda_{1b}\cos(k_y b/2)\cos(k_z c /2). 
\end{eqnarray}  
    Note that, under these constraints, $h_{21}(\mathbf{k})$ is guaranteed to be zero at the BZ boundary with $k_y = \pm \pi/b$.

Conversely, the lowest order term in $h_{22}(\mathbf{k})$ term is:
\begin{eqnarray}
 h_{22}(\mathbf{k})=\lambda_2   \sin(k_x a/2)\sin(k_y b/2),
\end{eqnarray} vanishing along the planes with $k_x=0$ and $k_y=0$, and having maximum amplitude at the PTBZ edges with $k_x=\pm \pi/a$ and $k_y = \pm \pi/b$.

Finally, following the constraint imposed by translations, the lowest order terms contributing to $h_{23}(\mathbf{k})$ are
\begin{eqnarray}
    h_{23}(\mathbf{k})=\lambda_{3a} \sin(k_x a/2)\sin(k_z c)\cos(k_y b/2)+\lambda_{3b}\sin(k_x a)\sin(k_z c/2)\cos(k_y b/2),
\end{eqnarray}
vanishing along the planes with $k_x=0$, $k_z=0$, and $k_y=\pm \pi/b$. Note that $h_{23}(\mathbf{k})$ is not generally zero at $k_x=\pm \pi/a$. 

The effect of finite SOC in the band structure is to lift band crossings, leading to the formation of small Fermi pockets at the Fermi surface. The $h_{22}(\mathbf{k})$ component is the only finite SOC term that is non-zero along the PTBZ boundary containing the $Y_{PT}$ point. Based on the electronic band splitting of approximately 10 meV along this line, we estimate the value of $\lambda_2$ to be approximately $5$ meV.

As the system preserves inversion and time-reversal symmetries, the four electronic bands emerge as two sets of  doubly degenerate bands for all $\mathbf{k}$ points with energies:
\begin{eqnarray}
    E^\pm_{SL+SOC}(\mathbf{k}) = h_{00}(\mathbf{k}) \pm \sqrt{h_{10}^2(\mathbf{k})+h_{30}^2(\mathbf{k})+h_{s}^2(\mathbf{k})},
\end{eqnarray}
where $h_{s}^2(\mathbf{k}) = \sum_i h_{2i}^2(\mathbf{k})$ corresponds to the contribution of SOC. Note that, as the different SOC components have different symmetries, $h_{s}^2(\mathbf{k})$ is not guaranteed to vanish along any line or plane of symmetry, and the two doubly-degenerate bands are never allowed to cross in the presence of SOC (except at the $Y_{PT}$ point).

\begin{figure}
\centering
\includegraphics[width=0.9\textwidth]{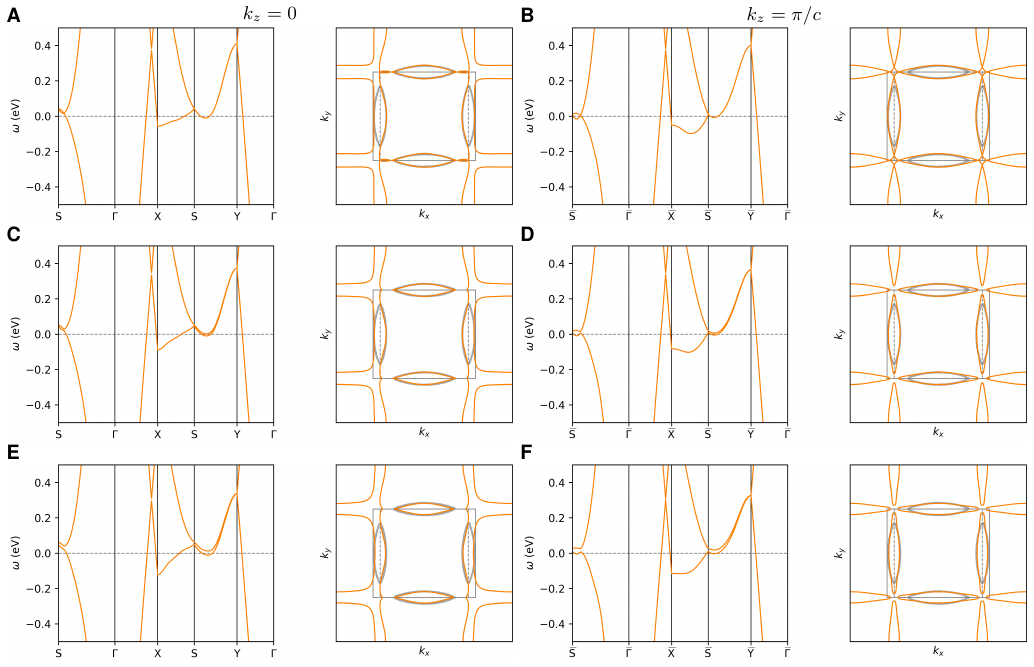}
  \caption{\textbf{Effect of SOC on the Fermi surface of the LTO phase.} Band structure and Fermi surfaces (full orange lines) of the LTO phase of La$_2$CuO$_4$ for $p=0.18$ in the $k_z=0$ (\textbf{A},\textbf{C},\textbf{E}) and  $k_z=\pi/c$ (\textbf{B},\textbf{D},\textbf{F}) planes, for different values of SOC. The dashed gray line marks the 2D pseudotetragonal BZ, and the full gray line marks the actual BZ in the $k_z=0$ plane. (\textbf{A}) and (\textbf{B}): no SOC, (\textbf{C})  and (\textbf{D}): nominal SOC, and (\textbf{E}) and (\textbf{F}): doubled SOC. 
  }
      \label{fig:fig_sm_lto}
\end{figure}

We plot representatively the Fermi surfaces for the LTO phase in the presence of SOC in Fig.~\ref{fig:fig_sm_lto}C-F. Note that, as expected from the symmetry-based argument above, the onset of SOC immediately lifts the band crossings around all the PTBZ faces, generally leading to the formation of small Fermi pockets for finite $k_z$. We overlay our results onto the Fermi surfaces reported by experiments in Fang et al. \cite{Fang2022}.

For Fig.~\ref{fig:properties}A in the main text, we use the tight-binding model in Eq.~\ref{Eq:TB_LTO}, with the parameters below it, taking all $k_z$-dependent terms to zero, which allows us to simplify the reciprocal space description in terms of a PTBZ. In the same spirit, we keep only the $h_{22}(\mathbf{k})$ SOC term, as this is the only term which does not break tetragonal symmetry and is the only SOC term that generates an anticrossing of the bands at the BZ edge with including $Y_{PT}$. We use an enhanced value of $\lambda_2=20$ meV and $\mu=-0.26$ eV.
We use $\mu=-0.26$ eV and an enhanced value of $\lambda_2=20$ meV to illustrate the sensitivity of the electronic structure to SOC, noting that interaction effects beyond DFT may further increase its effective magnitude.

\subsubsection*{Angle-resolved photoemission spectroscopy (ARPES) spectra}

ARPES measures the transition probability $\omega_{fi}$ of an optical excitation from a many-electron ground state to an excited state with one photoelectron. Theoretically, this transition probability can be estimated by Fermi's golden rule, assuming a weak light-matter interaction. Within the standard approximations, the ARPES signal can be well captured by 
\begin{eqnarray}
\omega_{fi} \approx \frac{2\pi}{\hbar} \sum_{\mathbf{k}} |M_{\mathbf{k}_f \mathbf{k}}|^2 A(\mathbf{k},\omega),
\end{eqnarray}
where $M_{\mathbf{k}_f \mathbf{k}}$ is the matrix element associated with the light-matter interaction between a Bloch state with momentum $\mathbf{k}$ and a photoelectron with momentum $\mathbf{k}_f$. $A (\mathbf{k},\omega)$ is the one-electron removal spectral function. For a detailed discussion, see \cite{Moser2017}. 

In the presence of multiple or structured internal degrees of freedom (here labelled by $\alpha$), the corresponding matrix element can be shown to be proportional to the Fourier transform of the Wannier or tight-binding orbitals:
\begin{eqnarray}
M_{\mathbf{k}_f\mathbf{k}}^\alpha \propto \langle \mathbf{k}_f|\mathbf{0},\alpha \rangle,
\end{eqnarray}
where $\alpha$ labels the eigenstates. To model the spectral contributions, we consider the non-interacting limit, assuming it is well captured by a Gaussian with intrinsic broadening $\sigma$:
\begin{eqnarray}
A(\mathbf{k},\omega-\epsilon_{\mathbf{k}}) \sim \frac{1}{\sqrt{2\pi}\sigma} e^{-(\omega-\epsilon_{\mathbf{k}})^2/(2\sigma^2)},
\end{eqnarray}
such that the ARPES spectra can be estimated as
\begin{eqnarray}
w_{fi} \propto \sum_\alpha |M^\alpha_{\mathbf{k}_f\mathbf{k}}|^2 A(\mathbf{k},\omega-\epsilon^\alpha_{\mathbf{k}}).
\end{eqnarray}

Below, we discuss in detail the implications of the matrix elements for the ARPES spectra in the HTT and LTO phases.

\begin{figure}
\centering
\includegraphics[width=0.9\textwidth]{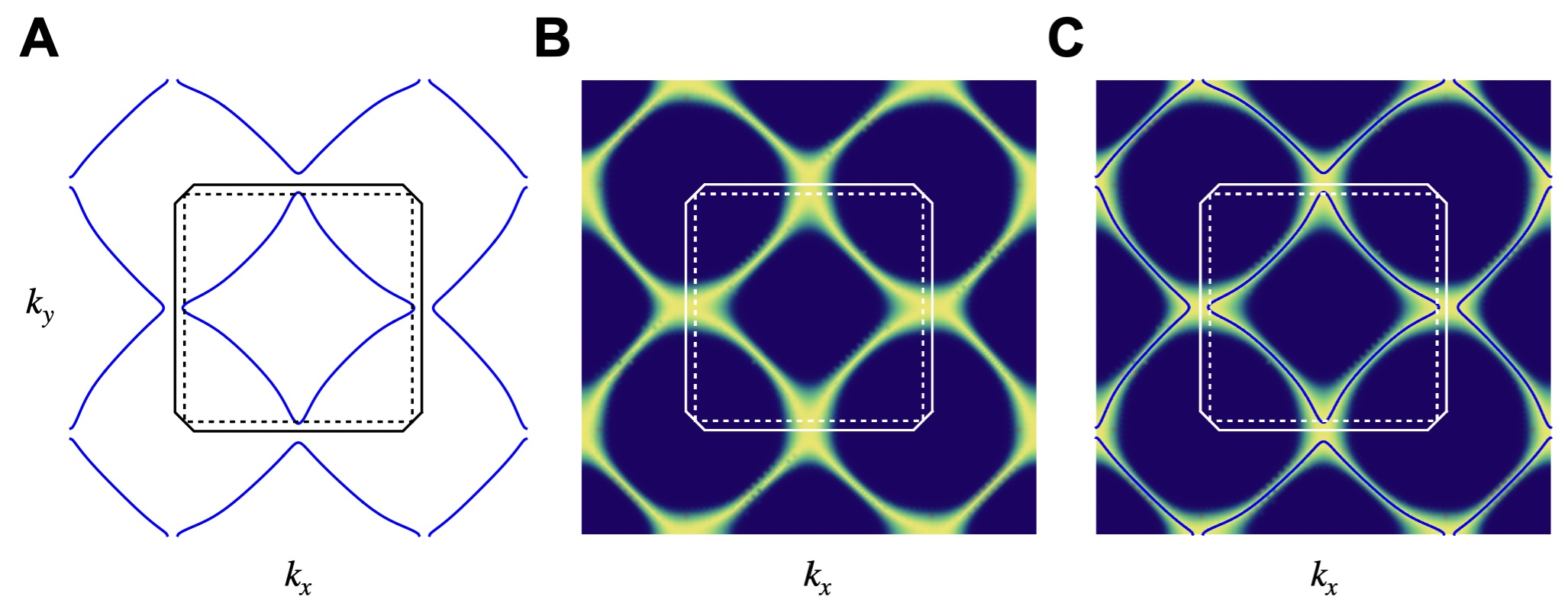}
  \caption{\textbf{ARPES spectra in the HTT phase.} 
  (\textbf{A}) Fermi surface of the HTT phase (blue lines) in an extended field of view in reciprocal space in the $k_z=0$ plane. We use the tight-binding model in Eq.~\ref{Eq:TB_HTT} with the parameters obtained from DFT listed below it, and $\mu=-0.2603$eV, corresponding to $p \approx 0.24$. The full black line marks the BZ edge at $k_z=0$ and the black dashed line marks the planar BZ.
  (\textbf{B}) ARPES spectra with $\sigma=0.1$. The BZs are marked as white lines. 
  (\textbf{C}) The two previous panels overlaid.
}
      \label{Fig:ARPES_HTT}
\end{figure}

\paragraph*{ARPES spectra in the HTT phase:} The tight-binding model for the HTT phase consists of one effective $d_{x^2-y^2}$ orbital. For normal emission, the matrix element can introduce a nodal structure to the ARPES maps, which is generally washed away for non-normal emission, as discussed in \cite{Moser2017}. Neglecting these factors, the ARPES spectra obtained purely from the spectral function in the HTT phase lead to a large Fermi surface as shown in Fig.~\ref{Fig:ARPES_HTT}, in agreement with observations in the overdoped regime outside the pseudogap phase.

For Fig.~\ref{fig:properties}E in the main text, we use the same parameters as in Fig.~\ref{fig:properties}B and $\sigma=0.025$ eV.

\paragraph*{ARPES spectra in the LTO phase with only in-plane terms and no SOC:} In the LTO phase, the crystal structure features a sublattice degree of freedom that gives rise to non-trivial matrix elements due to sublattice interference. The simplest model for the LTO phase consists of a two-dimensional model, for which all $k_z$-dependent terms are neglected, implying $h_3(\mathbf{k})=0$, in the absence of SOC. The Hamiltonian reads
\begin{eqnarray}
H_{SL-2D}(\mathbf{k}) &=& h_0(\mathbf{k})\hat{\tau}_0+h_1(\mathbf{k})\hat{\tau}_1,
\end{eqnarray}
with  eigenvalues
\begin{eqnarray}
E^\pm_{SL-2D}(\mathbf{k}) = h_0(\mathbf{k}) \pm |{h_1(\mathbf{k})}|,
\end{eqnarray}
and corresponding eigenvectors,
\begin{eqnarray}
\Psi^\pm_{SL-2D}(\mathbf{k}) = \frac{1}{\sqrt{2}}\left(1, \pm \frac{h_1(\mathbf{k})}{|h_1(\mathbf{k})|}\right),
\end{eqnarray}
corresponding to bonding ($-$) and anti-bonding ($+$) states. 

The tight-binding form of the Bloch wave function reads:
\begin{eqnarray}
|\Psi^\pm_{\mathbf{k}} \rangle_{SL-2D} =  \sum_{\mathbf{R}} e^{i\mathbf{k}\cdot \mathbf{R}} |\mathbf{R},\pm\rangle_{SL-2D},
\end{eqnarray}
where the sum ranges over all lattice sites $\mathbf{R}$, and
\begin{eqnarray}
 |\mathbf{R},\pm\rangle_{SL-2D} = \frac{1}{\sqrt{2}}\left(e^{i\mathbf{k}\cdot \mathbf{R}_A} |\mathbf{R}+\mathbf{R}_A, SL_A\rangle \pm   e^{i\mathbf{k}\cdot \mathbf{R}_B}  \frac{h_1(\mathbf{k})}{|h_1(\mathbf{k})|} |\mathbf{R}+\mathbf{R}_B, SL_B\rangle \right),
\end{eqnarray}
where $\mathbf{R}_{A,B}$ gives the position of the respective sublattice within the unit cell.

The photoemission matrix element can then be estimated as
\begin{eqnarray}
M^{{SL-2D}\pm}_{\mathbf{k}_f\mathbf{k}}& \propto& \langle \mathbf{k}_f | \mathbf{0}, \pm \rangle_{SL-2D}\propto \left(1 \pm \frac{h_1(\mathbf{k})}{|h_1(\mathbf{k})|}  \right) \langle \mathbf{k}_f | \mathbf{0}, SL_{A/B}\rangle,
\end{eqnarray}
where we used in-plane momentum conservation and assumed the orbitals in the two sublattices to be the same.

\begin{figure}
\centering
\includegraphics[width=0.9\textwidth]{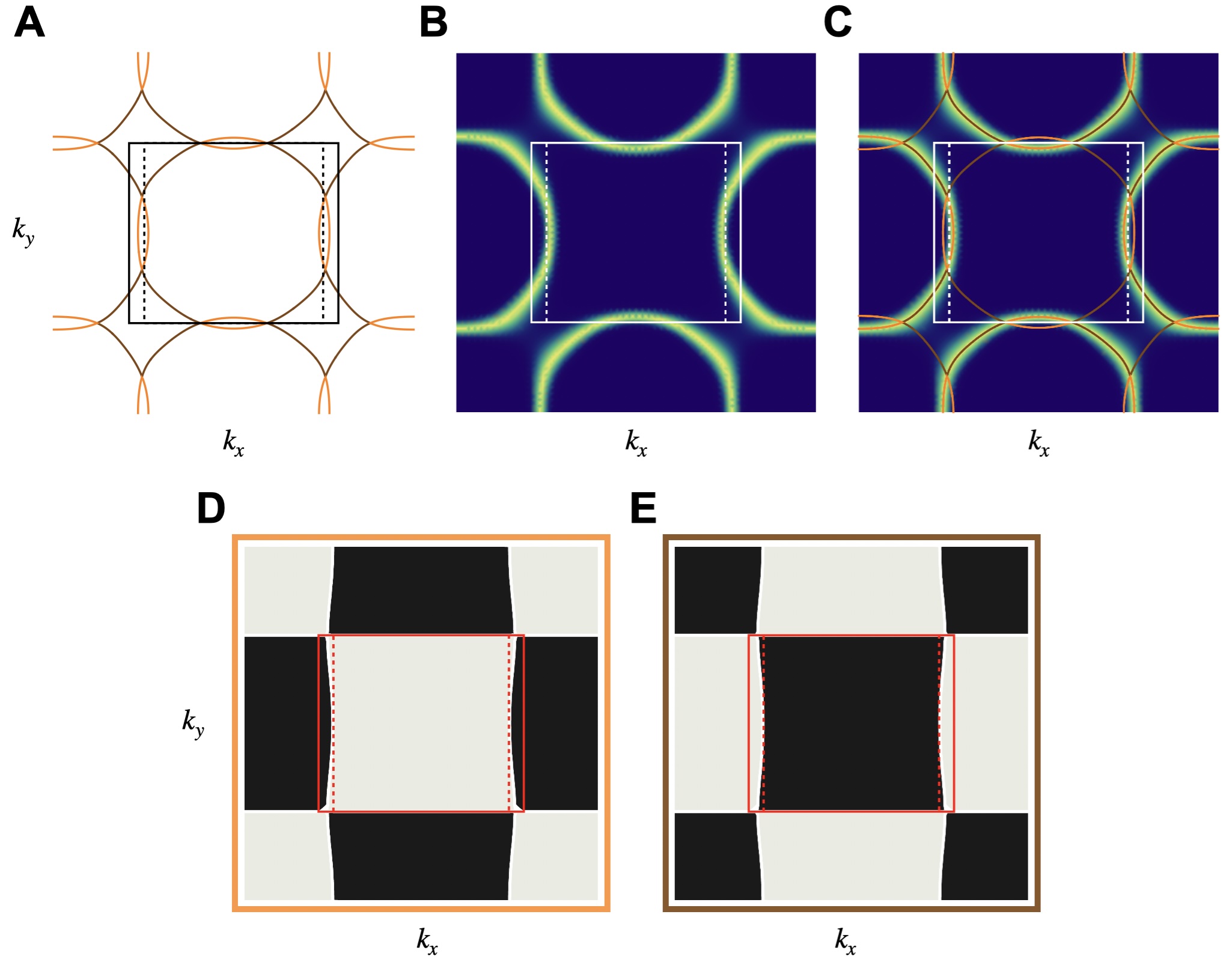}
  \caption{ \textbf{ARPES spectra in the LTO phase at zero doping with zero SOC.} 
  (\textbf{A}) Fermi surfaces of the LTO phase. The orange and brown lines correspond to the bonding and anti-bonding bands, respectively. We use the tight-binding model in Eq.~\ref{Eq:TB_LTO} with the parameters obtained from DFT listed below it, and $\mu=0$eV, corresponding to $p=0$.The full black line marks the BZ edge at $k_z=0$ and the black dashed line marks the PTBZ.
  (\textbf{B}) ARPES spectra with $\sigma=0.1$ showing one single large Fermi surface due to matrix element effects. 
  (\textbf{C}) The two previous panels overlaid. 
  (\textbf{D}) and (\textbf{E}) Normalized matrix elements for the bonding and anti-bonding bands, respectively. Light-gray corresponds to one, while black corresponds to zero. The BZs in panels (\textbf{B}-\textbf{E}) are displayed in different colors for better contrast.
  }
      \label{Fig:ARPES_LTO_0}
\end{figure}

\begin{figure}
\centering
\includegraphics[width=0.9\textwidth]{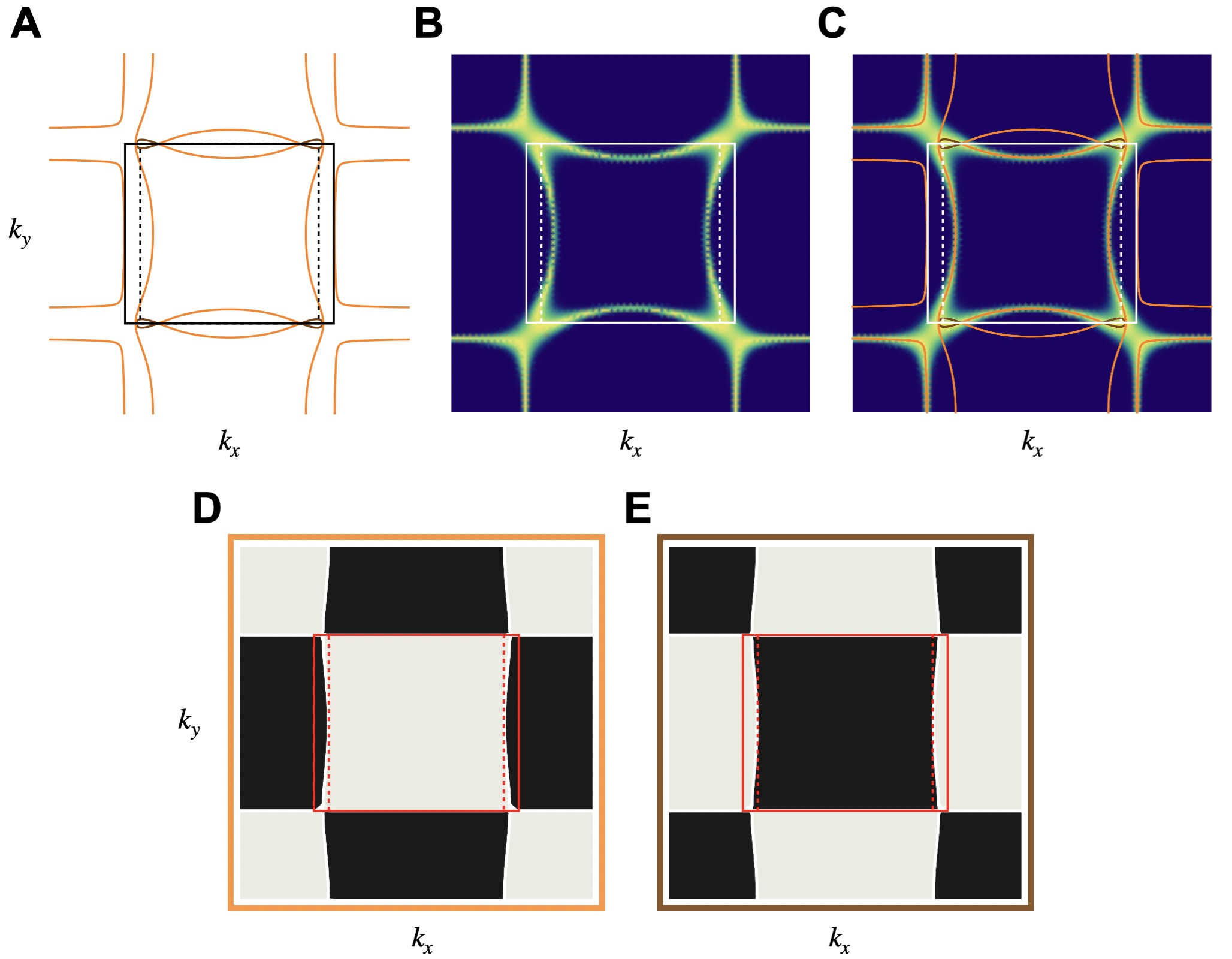}
  \caption{ \textbf{ARPES spectra in the LTO phase at finite doping with zero SOC.} Same as Fig.~\ref{Fig:ARPES_LTO_0}, but with  $\mu=-0.2694$eV, corresponding to $p\approx 0.18$. 
  }
      \label{Fig:ARPES_LTO_d}
\end{figure}

The matrix elements impose different form factors for the bonding and anti-bonding bands. In particular, as controlled by the $h_1(\mathbf{k})$ term associated with translations by fractional lattice vectors (see Eq. \ref{Eq:TB_LTO}), these form factors change as one crosses the first BZ. We display the expected ARPES spectrum for the two bands expected in LTO for zero doping in Fig.~\ref{Fig:ARPES_LTO_0} and finite doping in Fig.~\ref{Fig:ARPES_LTO_d}. Despite the presence of two electronic bands corresponding to bonding and anti-bonding states, the ARPES spectrum displays a single large Fermi surface, as the matrix elements selectively erase complementary parts of the bonding and anti-bonding Fermi surfaces.

\paragraph*{ARPES spectra in the LTO phase with out-of-plane terms and no SOC:} The Hamiltonian in this case reads:
\begin{eqnarray}
H_{SL-3D}(\mathbf{k}) &=& h_0(\mathbf{k})\hat{\tau}_0+h_1(\mathbf{k})\hat{\tau}_1 + h_3(\mathbf{k}) \hat{\tau}_3,
\end{eqnarray}
with  eigenvalues
\begin{eqnarray}
E^\pm_{SL-3D}(\mathbf{k}) = h_0(\mathbf{k}) \pm \sqrt{h_1^2(\mathbf{k})+h_3^2(\mathbf{k})},
\end{eqnarray}
and eigenvectors,
\begin{eqnarray}
\Psi^\pm_{SL-3D}(\mathbf{k}) = \frac{1}{\sqrt{2}}\left(1,  \frac{h_1(\mathbf{k})}{h_3(\mathbf{k})\pm\sqrt{h_1^2(\mathbf{k})+h_3^2(\mathbf{k})}}\right),
\end{eqnarray}
corresponding to bonding and anti-bonding states, which go back to the result of the previous subsection in the limit $h_3(\mathbf{k})\rightarrow 0$. 

The photoemission matrix element can then be estimated as
\begin{eqnarray}
M^{SL-3D\pm}_{\mathbf{k}_f\mathbf{k}}& \propto& \langle \mathbf{k}_f | \mathbf{0}, \pm \rangle_{SL-3D}\propto \left(1 + \frac{h_1(\mathbf{k})}{h_3(\mathbf{k})\pm\sqrt{h_1^2(\mathbf{k})+h_3^2(\mathbf{k})}} \right) \langle \mathbf{k}_f | \mathbf{0}, SL_{A/B}\rangle.
\end{eqnarray}

As the $k_z$-dependent terms are much smaller than the in-plane terms, these introduce only minor corrections to the form factors discussed above. 

\paragraph*{ARPES spectra in the LTO phase with out-of-plane terms and SOC:} As discussed above, the most general form of the Hamiltonian in the presence of SOC is:
\begin{eqnarray}
H_{SL+SOC}(\mathbf{k}) &=& \sum_{\{a,b\}=0}^3h_{ab}(\mathbf{k})\hat{\tau}_a\otimes \hat{\sigma}_b
\end{eqnarray}
where $\hat{\tau}_i$ and $\hat{\sigma}_i$ are Pauli matrices in sublattice and spin spaces, respectively, and $h_{ab}(\mathbf{k})$ are functions of momentum. As discussed above, in the presence of inversion and time-reversal symmetry (implemented as $i \hat{\hat{\tau}}_0 \otimes \hat{\sigma}_2$, $\mathbf{k} \rightarrow -\mathbf{k}$, followed by complex conjugation), the only six terms are symmetry-allowed in the Hamiltonian.

The eigenvalues of this Hamiltonian are doubly degenerate and equal to:
\begin{eqnarray}
E^\pm_{SL+SOC}(\mathbf{k}) = h_{00}(\mathbf{k}) \pm  |\mathbf{h}(\mathbf{k})|,
\end{eqnarray}
where $ |\mathbf{h}(\mathbf{k})| =\sqrt{h_{10}^2(\mathbf{k})+h_{30}^2(\mathbf{k})+h_{s}^2(\mathbf{k})}$, with $h_s^2(\mathbf{k}) = h_{21}^2(\mathbf{k})+h_{22}^2(\mathbf{k})+h_{23}^2(\mathbf{k})$, and eigenvectors
\begin{eqnarray}
\Psi_1^- &=&  \left(\frac{(|\mathbf{h}|-h_{30}) (h_{22}+i h_{21})}{{h_{10}}^2+{hs}^2} , -\frac{(|\mathbf{h}|-{h_{30}}) ({h_{10}}+i {h_{23}})}{{h_{10}}^2+{hs}^2} , 0 , 1 \right),\\
\Psi_2^- &=& \left(-\frac{(|\mathbf{h}|-{h_{30}}) ({h_{10}}-i {h_{23}})}{{h_{10}}^2+{hs}^2} , \frac{i (|\mathbf{h}|-{h_{30}}) ({h_{21}}+i {h_{22}})}{{h_{10}}^2+{hs}^2} , 1 , 0 \right),\\
\Psi_1^+ &=& \left(-\frac{i (|\mathbf{h}|+{h_{30}}) ({h_{21}}-i {h_{22}})}{{h_{10}}^2+\text{hs}^2} , \frac{(|\mathbf{h}|+{h_{30}}) ({h_{10}}+i {h_{23}})}{{h_{10}}^2+{hs}^2} , 0 , 1 \right),\\
\Psi_2^+ &=& \left(\frac{(|\mathbf{h}|+{h_{30}}) ({h_{10}}-i {h_{23}})}{{h_{10}}^2+{hs}^2} , \frac{(|\mathbf{h}|+{h_{30}}) ({h_{22}}-i {h_{21}})}{{h_{10}}^2+{hs}^2} , 1 , 0 \right),
\end{eqnarray}
where we have omitted the explicit $\mathbf{k}$-dependence for conciseness.

The photoemission matrix elements summed over the degenerate bands read:
\begin{eqnarray}
M_{\mathbf{k}_f,\mathbf{k}}^{SL+SOC+} &\propto& 1+ \frac{(|\mathbf{h}|+{h_{30}}) ({h_{10}}-{ih_{21}})}{{h_{10}}^2+{hs}^2},\\ \nonumber
M_{\mathbf{k}_f,\mathbf{k}}^{SL+SOC-} &\propto& 1- \frac{(|\mathbf{h}|-{h_{30}}) ({h_{10}}-{ih_{21}})}{{h_{10}}^2+{hs}^2}.
\end{eqnarray}

\begin{figure}
\centering
\includegraphics[width=0.9\textwidth]{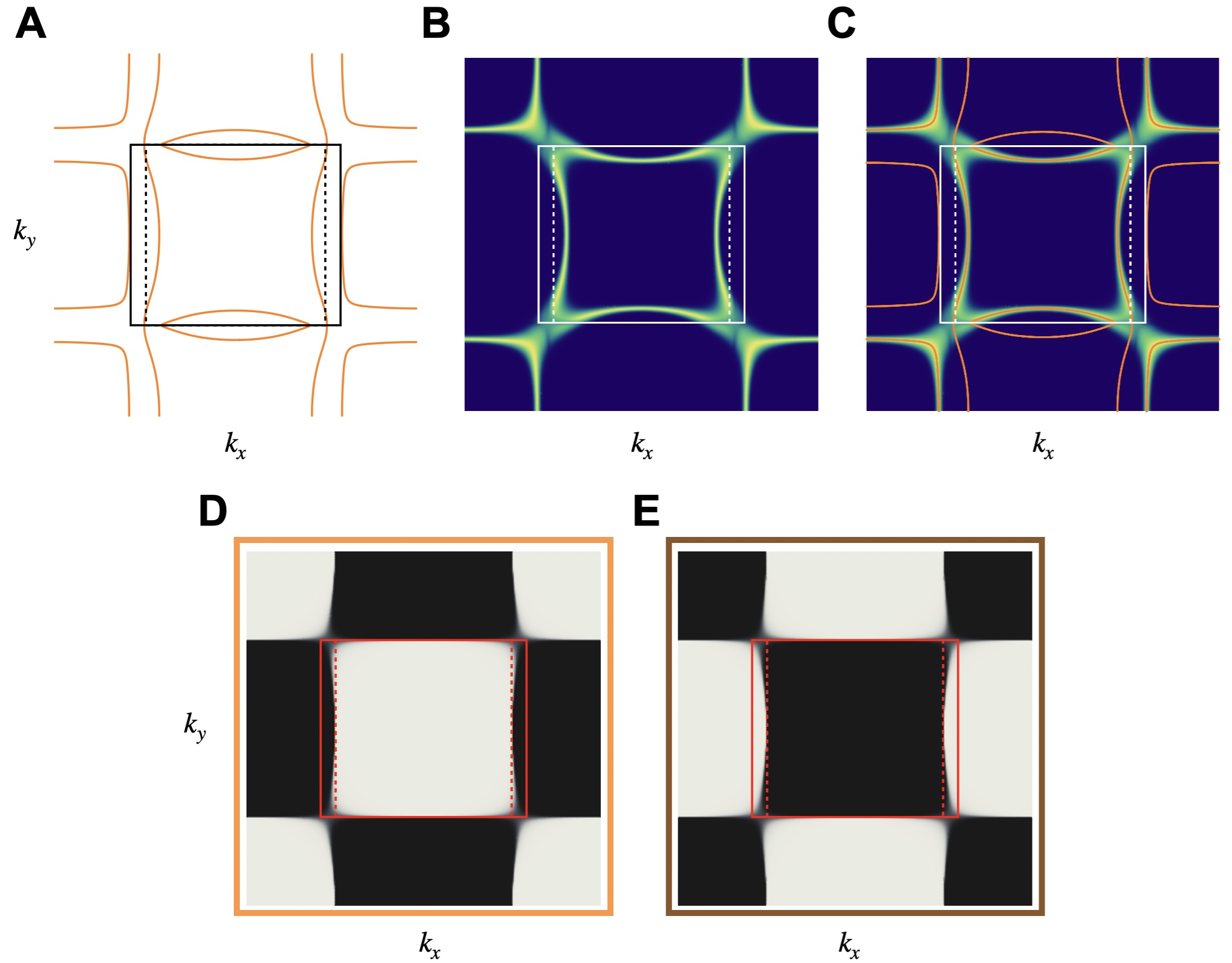}
  \caption{ \textbf{ARPES spectra in the LTO phase at finite doping and SOC.} 
  (\textbf{A}) Fermi surface of the LTO phase. In the presence of SOC the anti-bonding band is gapped and only the bonding band remains (orange line). We use the tight-binding model in Eq.~\ref{Eq:TB_LTO} with the parameters obtained from DFT listed below it and $\mu=-0.2873$ eV corresponding to $p\approx 0.18$. SOC is implemented through the term $h_{22}(\mathbf{k})$ with an enhanced $\lambda_2=0.02$ meV. The full black line marks the BZ edge at $k_z=0$ and the black dashed line marks the PTBZ.
  (\textbf{B}) ARPES spectra with $\sigma=0.05$ showing only parts of the Fermi surface due to matrix element effects. 
  (\textbf{C}) The two previous panels overlaid. 
  (\textbf{D}) and (\textbf{E}) Normalized matrix elements for the bonding and anti-bonding bands, respectively. Light-gray corresponds to one, while black corresponds to zero. The BZs in panels (\textbf{B}-\textbf{E}) are displayed in different colors for better contrast. }
      \label{Fig:ARPES_LTO_d_SOC}
\end{figure}

Note that these matrix elements go back to the results of the previous sections in the limit $h_{2i}(\mathbf{k})\rightarrow 0$ and $h_{30} (\mathbf{k})\rightarrow 0$. As the $h_{30}(\mathbf{k})$ and SOC terms are generally smaller than $h_{10}(\mathbf{k})$, they introduce only minor corrections to the form factors discussed above. For illustration, the Fermi surfaces and ARPES spectra in the presence of SOC are displayed in Figs.~\ref{Fig:ARPES_LTO_d_SOC}. 

For Fig.~\ref{fig:properties}D in the main text, we use the same parameters as in Fig.~\ref{fig:properties}A with $\sigma=0.025$ eV.

The explicit dependence of the spin DOF in these matrix elements could allow for spin-resolved ARPES. The spin-resolved matrix elements associated with each doubly-degenerate band are:
\begin{eqnarray}
M_{\mathbf{k}_f,\mathbf{k}}^{+\uparrow} &\propto& 1- \frac{(|\mathbf{h}|-{h_{30}}) ({h_{10}}-{ih_{21}} - {h_{22}}-i {h_{23}})}{{h_{10}}^2+{hs}^2},\\ \nonumber
M_{\mathbf{k}_f,\mathbf{k}}^{+\downarrow} &\propto& 1- \frac{(|\mathbf{h}|-{h_{30}}) ({h_{10}}-i {h_{21}+h_{22}}+i {h_{23}})}{{h_{10}}^2+{hs}^2},\\ \nonumber
M_{\mathbf{k}_f,\mathbf{k}}^{-\uparrow} &\propto& 1+\frac{(|\mathbf{h}|+{h_{30}}) ({h_{10}}-{ih_{21}} - {h_{22}}-i {h_{23}})}{{h_{10}}^2+{hs}^2},\\ \nonumber
M_{\mathbf{k}_f,\mathbf{k}}^{-\downarrow} &\propto& 1+ \frac{(|\mathbf{h}|+{h_{30}}) ({h_{10}}-{ih_{21}} + {h_{22}}+i {h_{23}})}{{h_{10}}^2+{hs}^2},
\end{eqnarray}
where we omitted the explicit $\mathbf{k}$-dependence for conciseness.

\newpage





\end{document}